\documentclass[aps,prc,twocolumn,showpacs,preprintnumbers,nofootinbib,float,superscriptaddress]{revtex4-1}
\usepackage{colortbl}
\usepackage{epsfig}
\usepackage[dvipsnames]{xcolor}
\usepackage{float,amsmath,amssymb}
\usepackage{graphicx}
\usepackage{url}
\usepackage{bbold}
\usepackage[utf8]{inputenc}
\usepackage{physics}
\usepackage[normalem]{ulem}
\usepackage{textcomp}
\usepackage[caption=false]{subfig}
\usepackage[labelfont=bf]{caption}
\usepackage{booktabs,multirow}

\DeclareCaptionLabelFormat{andtable}{#1~#2  \&  \tablename~\thetable}

\newcommand{\xpom}{x_{I\hspace{-0.3em}P}}

\newcommand{\mcal}{\mathcal}

\newcommand{\besk}{{\mathrm{K}}}

\newcommand{\dipole}{ {S}^{(2)} }
\newcommand{\tripole}{ {S}^{(3)} }

\newcommand{\aem}{\alpha_{\text{em}}}
\newcommand{\cf}{C_{\text{F}}}
\newcommand{\nlo}{{\textnormal{NLO}}}

\newcommand{\ov}[1]{\overline{#1}}

\newcommand{\rmd}{\mathrm{d}}
\newcommand{\xbj}{{x}}

\newcommand{\as}{{\alpha_{\mathrm{s}}}}
\newcommand{\rt}{{\mathbf{r}}}

\newcommand{\bt}{{\mathbf{b}}}
\newcommand{\Rt}{{\mathbf{R}}}

\newcommand{\Pt}{{\mathbf{P}}}
\newcommand{\Kt}{{\mathbf{K}}}

\newcommand{\xt}{{\mathbf{x}}}
\newcommand{\kt}{{\mathbf{k}}}

\newcommand{\Deltat}{\boldsymbol{\Delta}_\perp}

\newcommand{\nc}{N_\mathrm{c}}

\newcommand{\appropto}{\mathrel{\vcenter{
  \offinterlineskip\halign{\hfil$##$\cr
    \propto\cr\noalign{\kern2pt}\sim\cr\noalign{\kern-2pt}}}}}

\usepackage[breaklinks,colorlinks,citecolor=citcolor,urlcolor=blue,linkcolor=lcolor]{hyperref}
\definecolor{lcolor}{rgb}{0.5,0,0}
\definecolor{citcolor}{rgb}{0,0.3,0.0}
\usepackage[capitalise]{cleveref}
\usepackage{todonotes}

\begin{document}

\title{Diffractive deep inelastic scattering in the dipole picture: the $q\overline q g$ contribution in exact kinematics }

\author{Abhiram Kaushik}
\affiliation{Department of Physics, University of Jyv\"askyl\"a, P.O. Box 35, 40014 University of Jyväskylä, Finland}
\affiliation{Helsinki Institute of Physics, P.O. Box 64, 00014 University of Helsinki, Finland}

\author{Heikki M\"antysaari}
\affiliation{Department of Physics, University of Jyv\"askyl\"a, P.O. Box 35, 40014 University of Jyväskylä, Finland}
\affiliation{Helsinki Institute of Physics, P.O. Box 64, 00014 University of Helsinki, Finland}

\author{Jani Penttala}
\affiliation{Department of Physics and Astronomy, University of California, Los Angeles, CA 90095, USA}
\affiliation{Mani L. Bhaumik Institute for Theoretical Physics, University of California, Los Angeles, CA 90095, USA}

\begin{abstract}
We compute the $q\bar q g$ contribution to the diffractive structure functions in high-energy deep inelastic scattering. The obtained result corresponds to a finite part of the next-to-leading-order contribution to the diffractive cross section. Previous phenomenological applications have included this contribution only in the high-$Q^2$ or high-$M_X^2$ limits in the case of a soft gluon, and we numerically demonstrate that these existing estimates do not provide a good approximation for the full $q\bar q g$ contribution. Furthermore, we demonstrate that in addition to the soft gluon contribution, there is an equally important soft quark contribution to the diffractive structure functions at high $Q^2$.


\end{abstract}

\maketitle 

\section{Introduction}

High-energy scattering processes probe the structure of protons and nuclei at small momentum fraction $x$, where parton densities are very large and the emergence of non-linear saturation effects is predicted. However, as of today, no conclusive evidence of gluon saturation has been observed~\cite{Morreale:2021pnn}. Discovering saturation effects at currently achievable collider energies, and probing in detail the structure of protons and nuclei in this part of the phase space where non-linear phenomena dominate, is a major goal of the next-generation Electron-Ion Collider (EIC)~\cite{AbdulKhalek:2021gbh} at Brookhaven and the LHeC/FCC-he~\cite{LHeC:2020van} at CERN.

Diffractive deep inelastic scattering (DIS) with nuclear targets is expected to be an especially powerful probe of gluon saturation. First, in a diffractive event  at least two gluons are exchanged with the target, rendering the process proportional to the \emph{squared} gluon distribution function at lowest order. Furthermore, the cross section to diffractively produce a system with a given invariant mass $M_X^2$ is not sensitive to non-perturbative objects such as fragmentation functions or light-front wave functions (LCWF) of bound states required to describe hadronization, e.g., in the case of inclusive hadron  or exclusive vector meson production.  

A natural framework to describe QCD in the saturation region is provided by the Color Glass Condensate (CGC)~\cite{Iancu:2003xm,Garcia-Montero:2025hys}. A particular advantage is that it allows one to describe inclusive and diffractive observables in terms of the same degrees of freedom, and to resum multiple scattering effects that are important in the high-density domain. In the dipole picture~\cite{Mueller:1994jq}, one calculates  a virtual photon splitting into a partonic state, which then interacts with the target and produces a state with a given invariant mass. The available $\gamma+p\to X(M_X)+p$ data from HERA~\cite{H1:2012xlc} have been successfully described by many CGC calculations. These calculations take into account the leading $|q\bar q\rangle$ Fock state of the virtual photon, as well as the $|q\bar q g\rangle$ state in the high-$Q^2$ or high-$M_X^2$ limit~\cite{Kowalski:2008sa,Munier:2003zb,Marquet:2007nf}, and in some cases resum corrections enhanced by large $\log M_X^2$~\cite{Lappi:2023frf}.

The partial next-to-leading order (NLO) corrections accounted for by calculating the $|q\bar q g\rangle$ contribution in approximate kinematics can be sizeable~\cite{Kowalski:2008sa}, which calls for complete NLO calculations to match the precision achieved at HERA and expected at the EIC. This is particularly important when one aims to disentangle linear and non-linear dynamics in future nuclear diffractive structure function data (see e.g.~\cite{Armesto:2022mxy} in the context of inclusive structure functions). 

The virtual photon splitting into a partonic Fock state is described in terms of LCWFs, which can be computed in light cone perturbation theory. There has been rapid progress in recent years in computing LCWFs at NLO accuracy. From the diffractive DIS perspective, the most important developments include the calculation of the virtual photon wave function at NLO accuracy, first in the massless quark limit~\cite{Beuf:2016wdz,Beuf:2017bpd,Hanninen:2017ddy}, and later with heavy quarks included~\cite{Beuf:2021qqa,Beuf:2021srj,Beuf:2022ndu}. When coupled with an approximate version of the NLO Balitsky-Kovchegov (BK) energy evolution equation~\cite{Balitsky:2008zza,Lappi:2016fmu,Beuf:2014uia}, a good description of the inclusive structure function data has been obtained~\cite{Hanninen:2022gje,Casuga:2025etc}. The same LCWFs have also been applied to calculate the $|q\bar q g\rangle$ contribution to the diffractive structure function in exact eikonal kinematics\footnote{By exact eikonal kinematics we refer to the partonic  processes ($\gamma^*\to q\bar q$ and $q \to qg$) computed without any kinematical approximations. Eikonal approximation is applied to the interaction with the target shockwave.} in Ref.~\cite{Beuf:2022kyp}, and later the full diffractive cross section (in the massless quark limit) has been obtained in Ref.~\cite{Beuf:2024msh} (see also pioneering works~\cite{Boussarie:2014lxa,Boussarie:2016ogo,Boussarie:2019ero}). However, so far these results have not been applied to phenomenology, and the importance of the NLO corrections has not been quantified.

The purpose of this work is to take a step towards phenomenological calculations of diffractive cross sections in the CGC approach at NLO accuracy. In particular, we present the first numerical implementation of the $|q\bar q g\rangle$ contribution to the diffractive cross section, evaluated in exact eikonal kinematics. It is finite and computationally the most challenging part of the full NLO result obtained in Ref.~\cite{Beuf:2024msh}. The numerical challenge originates from the high-dimensional three-parton phase space integral, which requires integrating over the parton transverse coordinates separately in the amplitude and the conjugate amplitude, with the constraint that the invariant mass of the final state is fixed. The numerical implementation developed in this work can be applied to estimate the accuracy of the existing $|q\bar q g\rangle$ results, obtained in approximate kinematics and commonly used in phenomenology.
Moreover, it represents a crucial step towards our goal of confronting NLO CGC calculations with diffractive structure function data from HERA and the future EIC.

This manuscript is structured as follows. A finite subset of NLO corrections to the diffractive cross section for producing a $|q\bar qg\rangle$ system is reviewed in Sec.~\ref{sec:nlo_qqg}. Corresponding approximative results obtained in the high $Q^2$ or $M_X^2$ are discussed in Sec.~\ref{sec:limits}. Numerical results that quantify the accuracy of these approximative results are presented in Sec.~\ref{sec:numerics} before concluding in Sec.~\ref{sec:conclusions}. In Appendix~\ref{appendix:lnbeta} we furthermore discuss how the accuracy of the high-$Q^2$ limit can be improved by including a subset of contributions not enhanced by large $\log Q^2$.


\section{Production of $q\overline{q}g$  in exact kinematics}
\label{sec:nlo_qqg}

In Ref.~\cite{Beuf:2024msh}, the diffractive cross section at NLO is decomposed into three  contributions. The contribution where the $q\bar q g$ system interacts with the target shockwave is referred to as the ``trip'' contribution. This contribution, which is the focus of this work, is  finite and can be calculated separately. It is also expected to be the most computationally challenging part of the full NLO diffractive DIS cross section due to the high-dimensional phase space integral.
Previously, it has been studied only in approximate kinematics; these limits are discussed in Sec.~\ref{sec:limits}.
The two other contributions calculated in Ref.~\cite{Beuf:2024msh} are both UV divergent, and correspond to the cases where the $q\bar q$ dipole interacts with the shockwave and the final state can be either a $q\bar q$ or a $q\bar q g$ system (``dip''), and the interference between the $q\bar q$ and $q\bar q g$ contributions (``dip-trip''). These divergences cancel at the cross section level.

\begin{figure}
    \centering
    \includegraphics[width=0.45\columnwidth]{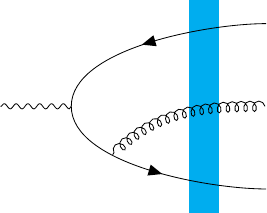}\quad
    \includegraphics[width=0.45\columnwidth]{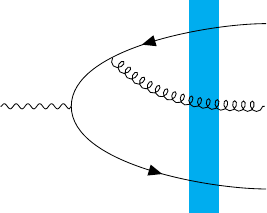}\\[2ex]
    \includegraphics[width=0.45\columnwidth]{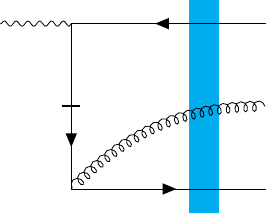}\quad
    \includegraphics[width=0.45\columnwidth]{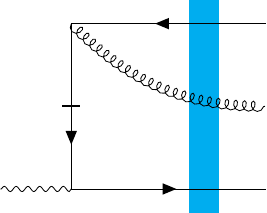}
\caption{
Contributions to the NLO-trip result for the diffractive cross section.  
}
\label{fig:tripdiags}
\end{figure}

The cross section for the ``trip'' contribution can be obtained from the diagrams shown in Fig.~\ref{fig:tripdiags}. Note that in light cone perturbation theory there are both regular and instantaneous contributions. This cross section has been first derived in Ref.~\cite{Beuf:2022kyp} and later reproduced as a part of the full NLO calculation~\cite{Beuf:2024msh}. It reads
\begin{multline}
\label{eq:trip}
         \biggl [ \frac{\dd[]{ \sigma^{\text{D}}_{\gamma^*_{\lambda} + A } }}{\dd[2]{\Deltat} \dd{M_X^2}}  \biggr ]_{\text{trip}} \!\!\!\!\!\!
      =   2 \pi\aem  \nc \sum e_f^2
      \int [\mathrm{dPS}]_{\text{trip}} \left (\frac{\as \cf}{2\pi} \right ) \\
    \times \mathcal{G}^{\nlo}_{\lambda,\text{trip}} 
    \left\langle 1-\tripole_{123}\right\rangle
    \left\langle 1-\tripole_{\ov 1 \ov 2 \ov 3}\right\rangle^*.
\end{multline}
Throughout this work we will refer to this contribution as the NLO-trip result.
Here $f$ refers to the quark flavor, and as the NLO cross section is derived in the massless limit, we include the three light quark flavors in the caclulation.
Detailed expressions for the hard factors $\mathcal{G}^{\nlo}_{\lambda,\text{trip}}$, where $\lambda=T,L$ refer to transverse and longitudinal photon polarization, are given in Ref.~\cite{Beuf:2024msh} and are not repeated here for brevity. 
The diffractive $\gamma^*+A$ cross section is related to the diffractive structure functions as
\begin{equation}
    \xpom F_{\lambda}^{\mathrm{D}(4)}(\xpom, Q^2, M_X^2, t) = \frac{Q^2}{4\pi^2 \aem} \frac{Q^2}{\beta}  \frac{\dd[]{ \sigma^{\text{D}}_{\gamma^*_{\lambda} + A } }}{\dd{t} \dd{M_X^2}}.
\end{equation}

The phase space integral in the mixed transverse coordinate--longitudinal momentum fraction space in Eq.~\eqref{eq:trip} is
\begin{multline}
\int [\mathrm{dPS}]_{\text{trip}} = \int \frac{
\dd[2]{\xt_{1}}\dd[2]{\xt_{2}}\dd[2]{\xt_{3}}
\dd[2]{\xt_{\ov 1}}\dd[2]{\xt_{\ov 2}}\dd[2]{\xt_{\ov 3}}
}{(2\pi)^6} \\
\times
\frac{e^{i \Deltat \vdot (\ov \bt -\bt)}}{(2\pi)^2}   
    \int_0^1  \dd[]{z_{1}} \dd[]{z_{2}} \dd[]{z_{3}} \delta(1-z_1 -z_2-z_3).
\label{eq:phase_space}
\end{multline}
Here $\xt_1,\xt_2$ and $\xt_3$ are the quark, antiquark and gluon transverse coordinates, and similarly $z_i$ denote the photon plus momentum fractions carried by these partons. The transverse momentum transfer $\Deltat$ is Fourier conjugate to the center-of-mass of the dipole $\bt = z_1 \xt_1 + z_2\xt_2 + z_3\xt_3$. The barred coordinates refer to those in the conjugate amplitude. The invariant $\xpom$ is the fraction of the target longitudinal momentum trasferred to the produced system in the infinite momentum frame:
\begin{equation}
    \xpom = \frac{M_X^2 + Q^2 - t}{W^2+Q^2-m_N^2}.
\end{equation}
Here $Q^2$ is the photon virtuality, $t\approx -\Deltat^2$ is the Mandelstam variable, and $W$ is the center-of-mass energy of the photon-nucleon system. Furthermore
\begin{equation}
    \beta = \frac{Q^2}{M_X^2+Q^2-t}.
\end{equation}

The eikonal propagation of the $q\bar q g$ system through the target color field is described in terms of Wilson line correlators:
\begin{equation}
\label{eq:tripole}
    \tripole_{123} = \frac{\nc}{2 C_F} \left[ \dipole_{13} \dipole_{23} - \frac{1}{\nc^2} \dipole_{12}\right].
\end{equation}
Here the dipole-target $S$-matrix reads
\begin{equation} 
\dipole_{12} = \frac{1}{\nc}  \tr V(\xt_1)V^\dagger(\xt_2) ,
\end{equation}
$\cf=(\nc^2-1)/(2\nc)$ and $V(\xt)$ is the Wilson line in the fundamental representation, which depends implicitly implicitly  on $\xpom$ through the JIMWLK equation~\cite{Mueller:2001uk}. 
In the cross section, Eq.~\eqref{eq:trip}, we need to average over target configurations, denoted by $\langle \mathcal{O}\rangle$, which is taken at the amplitude level for coherent diffraction that is the focus of this work.
Furthermore, we work in the mean field or large-$\nc$ limit, assuming that the average of Eq.~\eqref{eq:tripole} factorizes into a product of averages and drop terms suppressed by $1/\nc^2$.

In this work our goal is to present the first numerical evaluation of this $q\bar q g$ contribution to the diffractive cross section and not to perform detailed comparisons with the experimental data. As such, we adopt the following simple ($\xpom$-independent) parametrization for the tripole scattering amplitude $1-\tripole$:
\begin{equation}
\label{eq:N}
    1-\tripole_{123} = T(\bt)\left[1 - \tripole(\xt_{31},\xt_{32}) \right],
\end{equation}
where
\begin{equation}
    T(\bt)=\theta(R_A - |\bt|)
\end{equation}
and $\tripole(\xt_{31},\xt_{32})$ can be computed using \eqref{eq:tripole}.
Here $\xt_{ij} = \xt_i-\xt_j$, and we assume that the impact parameter dependence factorizes.
For the density profile, we adopt a simple step function (hard sphere with radius $R_A$).
For the $\bt$-independent dipole-target scattering amplitude we adopt the GBW model~\cite{Golec-Biernat:1998zce} 
\begin{equation}
    1-\dipole_{ij} = N(\xt_i,\xt_j) = 1 - e^{-\frac{\xt_{ij}^2 Q_{s}^2}{4}}.
\end{equation}

In our numerical implementation we choose $Q_{s}^2=0.4~\textrm{GeV}^2$. As the purpose of this work is to study the NLO-trip contribution and estimate the preision of the exisitng results obtained in different kinematical limits, we do not include small-$x$ evolution and consider a fixed value for the $Q_s^2$. As such, our results do not have an explicit dependence on $\xpom$.

With the assumption of a factorized impact parameter dependence, we can analytically integrate over $\Deltat$ and compute $\xpom F_\lambda^{D}(\xpom,Q^2,M_X^2) = \int_{-\infty}^0 \dd{t} \xpom F_\lambda^{D(4)}(\xpom,Q^2,M_X^2,t)$.  As the hard factors are independent of $\bt$ and $\overline{\bt}$, in this case the impact parameter integrals in Eq.~\eqref{eq:trip} can be simply calculated as
\begin{equation}
    \int \dd[2]{\Deltat} \int \dd[2]{\bt} \dd[2]{\overline{\bt}} \frac{e^{i\Deltat \cdot (\overline{\bt}-\bt)}}{(2\pi)^2} T(\bt)T(\overline \bt) = S_T,
    \label{eq:factorized_Delta_integral}
\end{equation}
where $S_T=\pi R_A^2$ is the target transverse size. In our numerical calculations we use $R_A = \sqrt{2B}$, where $B=4\,\mathrm{GeV}^{-2}$ is the diffractive slope for the proton.
Furthermore, we note that using a different density profile  instead of the hard sphere one in Eq.~\eqref{eq:N} would only affect this overall normalization factor.

\section{Production of $q\bar{q}g$ in approximative kinematics}
\label{sec:limits}

The cross section discussed in the previous section corresponds to diffractive $q\bar qg$ production in exact (eikonal) kinematics. The leading contributions to the $q\bar{q}g$ production at high $Q^2$ arise from kinematic configurations where one of the partons in the projectile wavefunction is soft (in the sense of the light-cone momentum fraction). Such configurations are known as ``aligned jet'' configurations. This is because the bulk of the virtual photon’s plus momentum is carried by one or more partons causing the resulting jet(s) to be aligned with the photon current. Specifically, it was shown in Ref.~\cite{Hauksson:2024bvv} that aligned jet configurations result in a strong scattering with the shockwave, and as such dominate the diffractive cross section 
in the leading twist approximation. However, it is important to clarify that the ``soft'' parton in these configurations is not arbitrarily soft, but rather has a light-cone momentum fraction of size $z\sim Q_s^2/Q^2\ll 1$. 

Such a result has long been established for the case of a soft gluon, with a tripole ($q\bar{q}g$) interacting with the shockwave. This contribution is known as the W\"usthoff  result (also sometimes referred to as the GBW result) and originally obtained in Refs.~\cite{Wusthoff:1997fz,GolecBiernat:1999qd,GolecBiernat:2001mm}. In addition to this, it was shown  in Ref.~\cite{Hauksson:2024bvv} that the soft quark (antiquark) limit should also give a large contribution to the cross section.

In addition to the  two cases discussed above, which correspond to the leading $\log Q^2$ contributions at large $Q^2$, there is an another important aligned jet configuration where the soft parton---in this case, the gluon---is indeed arbitrarily soft, with the gluon momentum fraction $z_3 \ll 1$ 
and the mass of the diffractively produced system very large, $M_X^2\gg Q^2$. This limit, sometimes referred to as the Munier--Shoshi result, has been considered by several authors \cite{Kowalski:2008sa,Munier:2003zb,Kovchegov:1999ji,Bartels:1999tn,Kopeliovich:1999am,Kovchegov:2001ni,Golec-Biernat:2005prq}. 
However, this limit also requires inclusion of the emission-after-shockwave contribution and therefore cannot be obtained directly from the NLO-trip contribution  \eqref{eq:trip}; instead it must be derived from the full NLO cross section~\cite{Beuf:2022kyp}. 
In principle, as the mass of the diffractive system gets very large, one has to also resum the emission of multiple soft gluons using the Kovchegov-Levin equation~\cite{Kovchegov:1999ji}. In this work we  compare the NLO-trip cross section to the ``Munier--Shoshi'' limit, in addition to the soft quark and gluon limits discussed above, due to the same final state.

In the following we briefly discuss the soft gluon aligned jet contribution to $q\bar{q}g$ production at high $Q^2$ in Sec.~\ref{sec:softg}. The soft quark contribution is then considered in Sec.~\ref{sec:softq}. As discussed in more detail later, in the soft quark production at high $Q^2$ the gluon emission can take place either before or after the shockwave, and the finite emission-before-shockwave contribution is calculated in this work. In Sec.~\ref{sec:MSlimit} we briefly review the $\beta\to 0$ limit.

\subsection{Soft gluon at high $Q^2$}
\label{sec:softg}



The leading power contribution to the diffractive structure function from the soft gluon aligned jet limit of the $q\bar{q}g$ system, i.e., the W\"usthoff result, has been well established and extensively and succesfully used in phenomenology~\cite{Marquet:2007nf,Kowalski:2008sa,Kugeratski:2005ck,Bendova:2020hkp,Lappi:2023frf}. 

 This contribution corresponds to the production of a hard $q\bar{q}$ dijet, recoiling against a semi-hard gluon jet. The latter is semi-hard in the sense that the transverse momentum of the gluon is of the order of the saturation scale. In coordinate space, this corresponds to the gluon being emitted far away from the $q\bar{q}$ dipole. As discussed in Section 3 of Ref.~\cite{Hauksson:2024bvv} (see also \cite{Iancu:2021rup, Iancu:2022lcw}), an essential condition for this configuration to result in strong scattering is that the gluon carries a lightcone momentum fraction $z_3\lesssim Q_s^2/Q^2$. Given these factors one could treat the $q\bar{q}g$ system as an adjoint dipole.
In the case of a large and uniform nucleus, the result given e.g. in Ref.~\cite{Kowalski:2008sa} can be written as
\begin{multline}
\label{eq:wusthoffqqbarg}
x_{\mathbb P}F_2^{D(3)}(\xbj, x_{\mathbb{P}}, Q^2) =\left(\sum e_f^2\right)\frac{\alpha_s \beta N_c C_F S_T}{4\pi^4} \\
\times \int_\beta^1\dd{x} P_{qg}\left(\frac{\beta}{x}\right)
 \int_0^{Q^2} \dd{k^2} k^4\log\left(\frac{Q^2}{k^2}\right)\,
 \\
 \times \left\{\int_0^\infty \dd{R} R \, J_2(\sqrt{1-x}k R)K_2(\sqrt{x}kR)\tilde{N}(R)\right\}^2,
\end{multline}
where the adjoint dipole at large $\nc$ reads $\tilde{N}(R) = 2N(R) - N(R)^2$. 
The splitting function is
\begin{equation}
    P_{qg}(z) = \frac{z^2+(1-z)^2}{2}.
\end{equation}
It has been explicitly shown in Ref.~\cite{Beuf:2022kyp} that this result can be derived from the NLO-trip contribution in the high-$Q^2$ limit.

\subsection{Soft quark at high $Q^2$}
\label{sec:softq}

To obtain the soft quark contribution to $q\bar{q}g$ production we closely follow the treatment in Ref.~\cite{Hauksson:2024bvv} where TMD factorization was established for diffractive jet production in photon-nucleus interactions. In the formalism developed therein, it was shown for diffractive processes that the soft parton in aligned jet configurations could be treated as a part of the target wavefunction, with a corresponding factorization established in terms of a diffractive TMD.

In Ref.~\cite{Hauksson:2024bvv} the soft quark contribution to the diffractive dijet production, as well as to the diffractive structure function, was calculated. As discussed in Sec.~\ref{sec:nlo_qqg}, the processes where the $q\bar q g$ system interacts with the shockwave form a finite subset of NLO corrections to the diffractive cross section. In this Section, we adapt the calculation of Ref.~\cite{Hauksson:2024bvv} to extract the emission-after-interaction contribution in the soft (anti)quark limit, without including the (divergent) emission-after-shockwave contribution. This is done by first expressing the LCWF in momentum space and then Fourier transforming it to coordinate space to describe interaction with the target. Finally, the result is transformed back to momentum space where the final-state kinematics can be specified.



\subsubsection{Amplitude for $q\bar{q}g$ production with a soft quark}

We calculate the diffractive production of a $|q\bar q g\rangle$ state at high $Q^2$, in the kinematics where the quark is soft, i.e. it carries a small fraction of the photon plus momentum: $z_1 \ll 1$. The transverse momentum of the quark can be of the order of the saturation scale: $|\kt_1| \sim Q_s$, which is why this quark is referred to as semi-hard. The contribution from the soft antiquark configuration can be obtained analogously. The discussion in this section closely follows that of Ref.~\cite{Hauksson:2024bvv}, where the reader is also referred to for futher details.

In the absence of scattering, the $q\bar{q}g$ component of the transverse virtual photon light-cone wavefunction has the following general form in momentum space,
\begin{align}
	\label{qqgmom}
	&\big|\gamma_{\scriptscriptstyle T}^{i}\big\rangle_{q\bar{q}g} 
	= \,t^a_{\alpha\beta}
	\int_{0}^{1}  \dd{z_1} \dd{z_2} \dd{z_3}
	\delta(1 - z_1 - z_2 -z_3)\nonumber\\
        &\times \int 
        \dd[2]{\kt_1} \dd[2]{\kt_2} \dd[2]{\kt_3} 
	\delta^{(2)}(\kt_1 + \kt_2 + \kt_3)
	\nonumber \\
	& \times
	\Psi^{im}_{\lambda_{1}\lambda_{2}}(z_1,\kt_1,
	z_2,\kt_2,
	z_3,\kt_3)\,\nonumber\\
	&\times \big|{q}_{\lambda_{1}}^{\alpha}(z_1, \kt_1)\,
	\bar{q}_{\lambda_{2}}^{\beta}(z_2, \kt_2)\,
	g_m^a(z_3, \kt_3)
	\big\rangle.
	\end{align}
Note that in this work we follow the conventions of Ref.~\cite{Hauksson:2024bvv} which differ slightly from those of Refs.~\cite{Beuf:2022kyp,Beuf:2024msh}.
Here $\Psi^{im}_{\lambda_{1}\lambda_{2}}$ is the $q\bar q g$ amplitude, and $z_j = k_j^+/q^+$ and $\kt_j$ are the light-cone momentum fraction and transverse momentum of the quark ($j=1$), antiquark ($j=2$) and the gluon ($j=3$). 
Furthermore $\lambda_1$ and $\lambda_2$ are the helicity indices of the quark and antiquark, and $\alpha$, $\beta$ and $a$ are the color indices of the quark, antiquark and gluon. 
Repeated indices are assumed to be summed over. 
The transverse photon and gluon polarization states are denoted by $i$ and $m$, corresponding to linear polarization states $i,m=1,2$. 
Note that, similarly as in the soft gluon case, we aim for extracting the $\log Q^2$ enhanced part of the leading twist contribution, which originates from the transverse photon.

We are in particular interested in the case where the gluon is emitted by the antiquark, since this is the case where our calculation that does not include the emission-after-interaction contribution differs from that of Ref.~\cite{Hauksson:2024bvv}. If the gluon is emitted by the quark which ends up being semi-hard in the final state, the emission-after-shockwave contribution is suppressed. In the following we discuss how this difference can be understood by considering the spatial configurations of the three parton states.

\begin{figure*}[t]
	\begin{center}
        \includegraphics[width=0.44\textwidth]{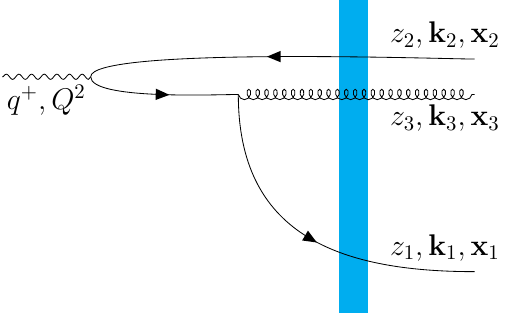}
        \hspace*{0.04\textwidth}
        \includegraphics[width=0.44\textwidth]{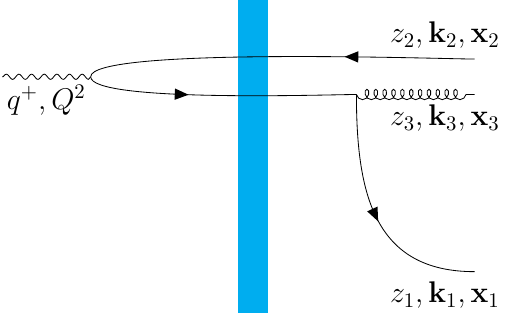}
	\end{center}
	\caption{
    Soft quark contribution to the diffractive cross section in the case where the quark emits the gluon before (left) or after (right) the shockwave. The quark is far away from the $\bar q g$ system.}
\label{fig:fromq}
\end{figure*}

Let us first consider the case where the quark emits the gluon. This contribution is shown in Fig.~\ref{fig:fromq}. The diagrams have been drawn in a way that makes the spatial separation of the partons manifest, and depict gluon emission before the shockwave (left panel) and after the shockwave (right panel). In addition to these, there is also a contribution from the emission of a gluon through an instantaneous interaction but we neglect this as it is power suppressed (see discussion in Appendix C of \cite{Hauksson:2024bvv}).
For the $q\bar{q}g$ system to be put on-shell, the partonic configuration interacting with the shockwave must have a large transverse size, $R \sim 1/Q_s$. This requirement ensures that color transparency is avoided.

Unless the longitudinal momentum is shared between the quark and the antiquark in a highly asymmetric manner, the initial $\gamma^* \to q' \bar{q}$ splitting is hard, i.e., the quark and antiquark are produced very close to each other but with a large relative transverse momentum $|\Pt| \gg Q_s$. Here $q'$ refers to the quark before the $q'\to qg$ splitting that takes place before the shockwave. If this subsequent splitting is such that the quark ends up with a small fraction of the longitudinal momentum of the $q'$, i.e., $z_1\sim Q_s^2/Q^2$, the quark is emitted at a distance $1/Q_s$ from the $\Bar{q}g$ system, ensuring a strong scattering with the target. The transverse momentum of this quark is of the order of the saturation scale.

If the $q'\to qg$ splitting takes place after the shockwave, the initial $q'\bar{q}$ pair is very small ($R\sim 1/Q)$ and this contribution vanishes due to color transparency.
Consequently, the contribution obtained in Ref.~\cite{Hauksson:2024bvv} exactly corresponds to the case that we are interested in this work, namely the one where the $|q\bar q g\rangle$ state interacts with the target. In particular, there is no need to subtract the post-shockwave emission when we consider gluon emission from the quark.

\begin{figure*}[t]
	\begin{center}
		\includegraphics[width=0.44\textwidth]{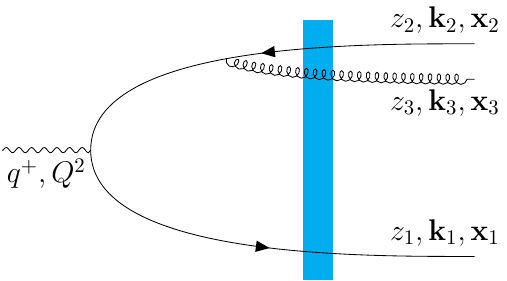}
        \hspace*{0.04\textwidth}
        \includegraphics[width=0.44\textwidth]{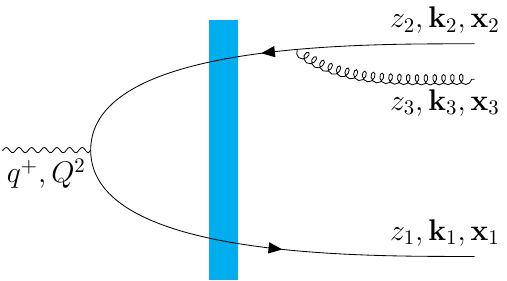}
	\end{center}
	\caption{
     Soft quark contribution to the diffractive cross section in the case where the antiquark emits the gluon before (left) or after (right) the shockwave. The quark is  far away from the $\bar q g$ system.
     }
\label{fig:fromqbar}
\end{figure*}

On the other hand, when the gluon is emitted by the antiquark, gluon emission after the shockwave is not suppressed by color transparency. In this case our calculation differs from that of Ref.~\cite{Hauksson:2024bvv}. 
The corresponding diagrams are shown in Fig.~\ref{fig:fromqbar}. As the quark carries a small longitudinal momentum fraction, the initial $\gamma^* \to q\bar{q}'$ splitting is soft, i.e. the quark is far away from the antiquark. Consequently there is strong scattering independently of whether the gluon is emitted by the antiquark before or after the shockwave. Since we want to compute the soft-quark component of the NLO ``trip'' contribution at large $Q^2$, we adapt the calculation of Ref.~\cite{Hauksson:2024bvv} to retain only the contribution of the left panel in Fig.~\ref{fig:fromqbar}. In the following, while we describe the essential components of this adapted calculation, we refer the reader to the aforementioned reference for the details.

The $q\bar{q}g$ amplitude where the gluon is emitted by the antiquark and before the shockwave is given by~\cite{Iancu:2022gpw}:
\begin{multline}
	\label{Psi1} 
	\Psi^{im}_{\lambda_1\lambda_{2}\,(\bar{q})}=
	\delta _{\lambda_{1}\lambda_{2}}\,
	\frac{1}{8q^+}\,
	\frac{e e_f g }{(2\pi)^6}\,
	\frac{1}{z_1(1-z_1)z_2z_3}\,
	\\
    \times \,\frac{1}{\sqrt{z_3}}\frac{\phi^{ij}(z_1,\lambda_1)\,k_1^j }
	{E_q+E_{\bar q'}-E_\gamma}\,
	\frac{\tau^{mn}(z_1,z_2,\lambda_1)P^n}
	{E_q+E_{\bar q}+E_g -E_\gamma}.
\end{multline}
The $(\bar{q})$ in the subscript indicates that the gluon is emitted by the antiquark.  The function $\phi^{ij}(z,\lambda)$ encodes the helicity structure of the photon splitting and is given by,
\begin{equation}
    \phi^{ij}(z,\lambda) \equiv (2z -1)\delta^{ij} + 2 \mathrm{i}\lambda\epsilon^{ij},
\end{equation}
where $\epsilon^{ij}$ is the Levi-Civita tensor in two dimensions. The function $\tau^{mn}(z_1, z_2, \lambda)$, which encodes the helicity structure of the $\bar{q}\to \bar{q}g$ vertex is given by,
\begin{align}
	\label{taudef}
	\tau^{mn}(z_1,z_2,\lambda)
	\equiv
	(1-z_1+z_2)
	\delta^{mn}
	+2\mathrm{i} \lambda z_3
	\varepsilon^{mn}.
\end{align}

The vector $\Pt$ (which appears in terms of its component of index $n$) corresponds to the relative transverse momentum of the antiquark and gluon,
\begin{align}
	\label{PandK_soft_quark}
	\Pt \equiv
	\frac{z_2 \kt_3 - z_3 \kt_2}
	{z_2 +z_3}
	\simeq 
	z_2 \kt_3 -
	z_3 \kt_2
	\,,\,\,\,
	\Kt \equiv \kt_2 + \kt_3.
\end{align}
Above, we have also defined $\Kt$ as the  momentum imbalance between the two hard jets in the final state, viz., the antiquark and the gluon. Since 
in coherent diffraction the momentum transfer from the target is  typically much smaller than other relevant momentum scales, the quark transverse momentum can be approximated as $\kt_1 \approx -\Kt$.

The two energy denominators in Eq. \eqref{Psi1} correspond to the photon splitting and the subsequent emission of a gluon by the antiquark. They are given by, 
\begin{align}
    \label{ED1}
    2q^+(E_q + E_{\bar{q}'} - E_\gamma) &=\frac{\kt_1^2}{z_1} + \frac{\kt_1^2}{1-z_1} + Q^2\nonumber\\
    &=\frac{\kt_1^2+z_1(1-z_1)Q^2}{z_1(1-z_1)}\,
\end{align}
and
\begin{align}
	\label{ED2}
	2q^+\big(E_q + E_{\bar q} + E_g -E_\gamma\big)&\,=
	\frac{\kt_1^2}{z_1}+
	\frac{\kt_2^2}{z_2}+
	\frac{\kt_3^2}{z_3}+
	Q^2 
	\nonumber\\ &= 
	\frac{\kt_1^2}{z_1(1-z_1)}+
	\frac{1-z_1}{z_2z_3}\,{\Pt^2}+
	Q^2\,
	\nonumber \\
	&\,
	\simeq
	\frac{\kt_1^2+\mcal{M}^2}{z_1},
\end{align}
where we have defined
\begin{equation}
    \mcal{M}^2 \equiv z_1 \left(\frac{\Pt^2}{z_2z_3} + Q^2\right)
\end{equation}
and approximated $1-z_1 \approx 1$.

Using these expressions for the denominators and taking the limit $z_1\to0$ wherever possible, the amplitude in Eq.~\eqref{Psi1} becomes,
\begin{align}
    \Psi^{im\,(1)}_{\lambda_1\lambda_2\,(\bar{q})}&=
	\delta _{\lambda_{1}\lambda_2}\,
    \frac{e e_f g q^+ }{2(2\pi)^6}\,
    \frac{z_1}{z_2 z_3\sqrt{z_3}}\nonumber\\
    &\times\,    \frac{\phi^{ij}(0,\lambda_1)k_1^j\,\tau^{mn}(0,z_2,\lambda_1)P^n }
	{(\kt_1^2+\mcal{M}^2)(\kt_1^2+z_1 Q^2)}\,.
\end{align}
The two transverse momentum factors in the denominator can be decomposed as follows:
\begin{align}
    \frac{1}{(\kt_1^2+\mcal{M}^2)(\kt_1^2+z_1 Q^2)} = \frac{A}{(\kt_1^2+\mcal{M}^2)} + \frac{B}{(\kt_1^2+z_1 Q^2)}\,
\end{align}
where,
\begin{align}
    A = -B = \frac{1}{z_1Q^2 - \mcal{M}^2} = -\frac{z_2 z_3}{z_1 \Pt^2}\,.
\end{align}
With this, the amplitude becomes
\begin{align}
    \Psi^{im\,(1)}_{\lambda_1\lambda_2\,(\bar{q})}&=
	\delta _{\lambda_{1}\lambda_2}\,
    \frac{e e_f g q^+ }{2(2\pi)^6}\,
    \frac{1}{\sqrt{z_3}}\,
    \phi^{ij}(0,\lambda_1)k_1^j\nonumber\\
    &\times\left(\frac{1}{\kt_1^2+\mcal{M}^2} - \frac{1}{\kt_1^2+z_1 Q^2}\right)\frac{\tau^{mn}(0,z_2,\lambda_1) P^n}{\Pt^2}\,.
\end{align}
Comparing this to Eq. (3.25) of Ref.~\cite{Hauksson:2024bvv}, we see that the overall structure is the same. The additional term $-\frac{1}{\kt_1^2+z_1 Q^2}$ above can be seen as a  subtraction of the emission-after-shockwave contribution.

The eikonal propagation of quarks and gluons through the shockwave is described by Wilson lines in fundamental or adjoint representation. Denoting the quark, antiquark and gluon transverse coordinates again by $\xt_1,\xt_2$ and $\xt_3$, the scattering can be described by performing a replacement~\cite{Hauksson:2024bvv}
\begin{multline}
\label{eq:qqg_wline}
    t^a_{\alpha\beta} \to [U^{ab}(\xt_3)V(\xt_1)t^bV^\dagger(\xt_2) - t^a]_{\alpha\beta} \\
    \approx [V(\xt_1)V^\dagger(\xt_1-\Rt)t^a-t^a]_{\alpha \beta}
\end{multline}
in Eq.~\eqref{qqgmom} Fourier transformed into the coordinate space.
Here $V$ and $U$ are the fundamental and adjoint Wilson lines, and we have defined $\Rt$ such that $\xt_3\approx \xt_2 = \xt_1-\Rt$. This effectively corresponds to a scattering of a quark-antiquark dipole. 
The Wilson lines depend on target configurations. In this work we focus on coherent diffraction, and as such we perform the average over these configurations at the level of the scattering amplitude. As a result of this average, the Wilson line structure in \eqref{eq:qqg_wline} becomes $-N(\Rt) t^a_{\alpha,\beta}$ when assuming a large uniform target.

Overall, the application of this three step procedure (Fourier transform to coordinate space, replacement $t^a_{\alpha,\beta}\to -N(\Rt)t^a_{\alpha,\beta}$, transform back to momentum space) can be done by performing the following replacements~\cite{Hauksson:2024bvv}
\begin{align}
\label{eq:momspacereplacement}
    \frac{k_1^j}{\kt_1^2+\mcal{M}^2}&\to\frac{K^j}{|\Kt|}\frac{\mcal{Q}_T(\mcal{M},\Kt,Y_\mathbb{P})}{\mcal{M}},\nonumber\\  \frac{k_1^j}{\kt_1^2+z_1Q^2}&\to\frac{K^j}{|\Kt|}\frac{\mcal{Q}_T(\sqrt{z_1}Q,\Kt,Y_\mathbb{P})}{\sqrt{z_1}Q}\,.
\end{align}
Here we have defined
\begin{equation}
    \mathcal{Q}_T(\mu, \Kt, Y_\mathbb{P}) = \mu^2 \int \dd{R} R J_1(|\Kt| R) K_1(\mu R) N(R).
\end{equation}
Here the dipole amplitude $N$ depends implicitly on $Y_\mathbb{P}$ or $\xpom$ as  discussed in Sec.~\ref{sec:nlo_qqg}. 

Using Eq.~\eqref{eq:momspacereplacement},  the final expression for the amplitude to  diffractively produce of a $q\bar{q}g$  state via the channel $\bar{q}'\to \bar{q}g$, such that the quark jet is semi-hard and the gluon is emitted before the shockwave, can be obtained:
\begin{multline}
    \label{final_qb_emission_amplitude}
	\Psi^{im,D}_{\lambda_1\lambda_2 (\bar{q})}
	\simeq \,\delta _{\lambda_{1}\lambda_2}\,
	\frac{e e_f g q^+ }{2(2\pi)^6}\,
	\frac{\Phi^{ijmn}_{(\bar{q})}(z_2,\lambda_1)}{\sqrt{z_3}}\,
	\frac{P^n}{\Pt^2}\,\frac{K^j}{|\Kt|} \\
    \times \left(\frac{\mcal{Q}_{T}(\mcal{M}, \Kt,Y_\mathbb{P})}{\mcal{M}} - \frac{\mcal{Q}_{T}(\sqrt{z_1}Q, \Kt,Y_\mathbb{P})}{\sqrt{z_1}Q}\right)\,.
\end{multline}
Here, the spinorial and polarisation structure is encoded in
\begin{align}
	\label{Phiqbar}
	\Phi^{ijmn}_{(\bar{q})}(z,\lambda)\,
	&\equiv\,
	-\,\phi^{ij}(0,\lambda)
	\tau^{mn}(0,z,\lambda)\nonumber\\
	&= 
	\big[\delta^{ij} -
	2\mathrm{i}\lambda\varepsilon^{ij}\big]
	\big[(1+z)\delta^{mn}
	+2\mathrm{i}\lambda (1-z)\varepsilon^{mn}\big].
\end{align}

It is instructive to compare Eq.~\eqref{final_qb_emission_amplitude} with the corresponding expression Eq.~(3.34) in Ref.~\cite{Hauksson:2024bvv} which includes gluon emissions both before and after the shockwave.  The two expressions differ in the soft factor with Eq.~\eqref{final_qb_emission_amplitude} having a linear combination of the $\mathcal{Q}_T$ terms instead of just $\mcal{Q}_T(\mcal{M},\Kt,Y_\mathbb{P})/\mcal{M}$.
As such, the second term with a minus sign can be seen to subtract the emission-after-shockwave contribution.
As we will explicitly see, without this subtraction, the amplitude $\Psi^{im,D}_{\lambda_1\lambda_2 (\bar{q})}$ would result in a divergent contribution to the structure function when squared and appropriately integrated over the hard transverse momentum $\Pt$ and other kinematic variables. This would correspond to the soft divergence from the final state gluon emission discussed in detail in  Ref.~\cite{Beuf:2024msh}. In our case, this divergence, which can be traced to the $1/|\Pt|$ dependence in the hard factor, is regulated by the fact that the soft factor vanishes when $|\Pt|\to0$ since $\mcal{M} \to \sqrt{z_1}Q$ in this limit. 

For completeness we also write down the corresponding amplitude $\Psi^{im,D}_{\lambda_1\lambda_2 (q)}$ for the process in which the quark emits the gluon. As noted earlier, we can directly use the result from Ref.~\cite{Hauksson:2024bvv} since  gluon emission after the shockwave does not contribute:
\begin{align}
	\label{q_emission_amplitude}
	\Psi^{im,D}_{\lambda_1\lambda_2(q)}
	&\simeq 
	\delta _{\lambda_{1}\lambda_2}\,
	\frac{e e_f g q^+ }{2(2\pi)^6}\,
	\frac{\Phi^{ijmn}_{(q)}(z_2,\lambda_1)}{z_3^{3/2}}\nonumber\\
    &\times	\frac{P^j}
	{\Pt^2+\tilde{Q}^2}\,
	\frac{K^n}
	{|\Kt|}\,
	\frac{\mcal{Q}_{T}(\mcal{M}, \Kt,Y_\mathbb{P})}{\mcal{M}}.
\end{align}
Here the spinorial and polarisation structure reads
\begin{align}
	\label{Phiq}
	\Phi^{ijmn}_{(q)}(z,\lambda)\,
	&\equiv\,
	-\,\phi^{ij*}(z,\lambda)
	\tau^{mn*}(z,0,\lambda)\nonumber \\
	&\hspace{-0.5cm}=
	(1-z)\big[(1- 2z ) \delta^{ij} +
	2i\lambda\varepsilon^{ij}\big]
	\big[\delta^{mn}
	-2\mathrm{i} \lambda \varepsilon^{mn}\big],
\end{align}
and $\tilde{Q}^2 \equiv z_2z_3Q^2$.

\subsubsection{Cross section for diffractive $q\bar{q}g$ production with a soft quark}

The cross section for diffractive $q\bar{q}g$ production with a soft quark is obtained by summing and squaring the amplitudes in Eqs.~\eqref{final_qb_emission_amplitude} and \eqref{q_emission_amplitude}, applying the appropriate normalization factor, and summing over the spin and polarization indices. This procedure is not affected by our restriction to pre-shockwave gluon emission, and we can directly use the result from Ref.~\cite{Hauksson:2024bvv}. The cross section can be written as a sum of three terms: the direct term from gluon emission by antiquark, the direct term from gluon emission by quark, and their interference. The direct contribution from gluon emission by the antiquark is given by 
\begin{multline}
	\label{cross1}
	\frac{\rmd\sigma_{(\bar{q}\bar{q})}^{\gamma_{\scriptscriptstyle T}^* A\rightarrow (q)\bar q g A}}
	{\rmd z_1 
	\rmd z_2
	\rmd z_3 \,
	\rmd^{2}\Pt
  	\rmd^{2}\Kt}
   = 
   \left( \sum e_f^2\right)\frac{S_{\perp} \alpha_{\rm em} N_c}{2\pi^4} \\
   \times 
   \delta_{z}   \frac{\alpha_s C_F}{P_{\perp}^2}
   \frac{1+(1-z_3)^2}{2z_3} \\
   \times  \left(\frac{\mcal{Q}_{T}(\mcal{M}, \Kt, Y_\mathbb{P})}{\mcal{M}} - \frac{\mcal{Q}_{T}(\sqrt{z_1}Q, \Kt, Y_\mathbb{P})}{\sqrt{z_1}Q}\right)^2.
   \end{multline}
We remind the reader that we work in the case where the target is a large uniform nucleus, and as such momentum transfer to the target is small.
A few comments on the notation used here: The ``$(q)$" in the superscript indicates that the quark is the soft parton in the final state. In the subscript there are two parton labels within parentheses.  The first label indicates the parton emitting the gluon in the amplitude and the second label indicates the parton emitting the gluon in the complex conjugate. In this case, ``$(\bar{q}\bar{q})$" indicates that this is the direct term from gluon emission by the antiquark. We have also introduced a shorthand notation for the delta function $\delta_z\equiv\delta(1-z_1-z_2 - z_3)\simeq\delta(1 - z_2 - z_3)$. 

The interference term can be written as,
\begin{multline}
	\label{cross3}
	\frac{\rmd\sigma_{(\bar{q}q)}^
	{\gamma_{\scriptscriptstyle T}^* A\rightarrow (q)\bar q g A}}
	{\rmd z_1 
	\rmd z_2
	\rmd z_3 \,
	\rmd^{2}\Pt
  	\rmd^{2}\Kt}
   = -
   \left(\sum e_f^2\right)\frac{S_{\perp} \alpha_{\rm em} N_c}{2\pi^4} 
   \delta_{z} \\
   \times 
   \frac{\alpha_s C_F}{P_{\perp}^2 + \tilde{Q}^2}\,
   \frac{z_2^2}{z_3} 
    \frac{\big|\mcal{Q}_{T}(\mcal{M}, \Kt,Y_{\mathbb P})\big|}{\mcal{M}}  \\
    \times \left(\frac{\mcal{Q}_{T}(\mcal{M}, \Kt, Y_\mathbb{P})}{\mcal{M}} - \frac{\mcal{Q}_{T}(\sqrt{z_1}Q, \Kt, Y_\mathbb{P})}{\sqrt{z_1}Q}\right).
\end{multline}

For completeness, we also reproduce the direct term from gluon emission by the quark which is unchanged from Ref.~\cite{Hauksson:2024bvv}:
\begin{align}
	\label{cross2}
	\hspace*{-0.6cm}
	\frac{\rmd\sigma_{(qq)}^
	{\gamma_{\scriptscriptstyle T}^* A \rightarrow (q)\bar q g A}}
	{\rmd z_1 
	\rmd z_2
	\rmd z_3 \,
	\rmd^{2}\Pt
  	\rmd^{2}\Kt}
   &= 
   \frac{S_{\perp} \alpha_{\rm em} N_c}{2\pi^4} 
   \left( \sum e_f^2\right)
   \delta_{z}\,
   \frac{\alpha_s C_F P_{\perp}^2}{(P_{\perp}^2 + \tilde{Q}^2)^2}\nonumber\\
   &\times    \frac{z_2^2 +(1-z_2)^2}{2z_3}\,
   \frac{\big|\mcal{Q}_{T}(\mcal{M}, \Kt,Y_{\mathbb P})\big|^2}{\mcal{M}^2}.
\end{align}

While this is not strictly relevant to the current work, we note that the linear combination of the two $\mcal{Q}_T$ terms in \eqref{cross1} and \eqref{cross3} spoils the possibility of TMD factorization of the diffractive cross section that is demonstrated in Ref.~\cite{Hauksson:2024bvv}. This is because the three different contributions to the differential cross section in Eqs. \eqref{cross1}, \eqref{cross3} and \eqref{cross2}, all have a different soft factor. However this is not too surprising as one would expect that any factorization framework would treat emissions before and after the shockwave on an equal footing. One can nevertheless follow --- to the extent possible --- the procedure in Sections 4 and 5 of Ref.~\cite{Hauksson:2024bvv}, to obtain the leading power contribution in our case.
Including the emission after the shockwave contributions would restore the TMD factorization~\cite{Hauksson:2024bvv}.

\subsubsection{Soft quark contribution to diffractive SIDIS and the structure function}
We have so far worked in the {\it projectile} picture where an intial color dipole radiates a gluon before scattering with the target shockwave. Now we shift to the {\it target} picture where this process can be viewed as a hard scattering off the virtual photon of a partonic constituent of a Pomeron in the target. Now, as  mentioned in the previous subsection, factorization in the target picture is not valid for the quantity we want to calculate, as we are not including the final state emissions. Nevertheless, this helps us reuse the machinery employed in Ref.~\cite{Hauksson:2024bvv} which greatly simplifies our derivation. 

In the target picture the soft quark is viewed as a constituent of the Pomeron. The Pomeron fluctuates into a quark-antiquark pair with the $s$-channel quark being on-shell and part of the final state, whereas the $t$-channel antiquark scatters with the virtual photon in a $2 \to 2$ process that produces a hard $\bar{q}g$ dijet in the final state. This is illustrated in Fig.~10 of Ref.~\cite{Hauksson:2024bvv}. 

The key step in moving to the target picture is to perform a change of variables, from the projectile longitudinal momentum fraction $z_1$ to  $x$. The latter is the `minus' momentum fraction of the $t$-channel antiquark with respect to the Pomeron. This allows us to write Eqs.~\eqref{cross1}, \eqref{cross3} and \eqref{cross2} as
\begin{widetext}
\begin{align}
    \label{cross1_tmd}
    \frac{\rmd\sigma_{(\bar{q}\bar{q})}^
	{\gamma_{\scriptscriptstyle T}^* A\rightarrow (q)\bar q g A}}
	{\rmd z_2
	\rmd z_3
	\rmd^{2}\Pt
  	\rmd^{2}\Kt\,
  	 \rmd \ln(1/x)}
   &= 
   \frac{1}{(1-x)}\frac{1}{Q^2}\,\frac{S_{\perp} \alpha_{\rm em} N_c}{2\pi^2} 
   \left( \sum e_f^2\right)\,\mcal{H}^{(\bar{q}\bar{q})}_T
 	(z_2,z_3,P_{\perp}^2,\tilde{Q}^2)\,\nonumber\\
   &\times \mcal{M}^2\left(\frac{\mcal{Q}_{T}(\mcal{M}, \Kt, Y_\mathbb{P})}{\mcal{M}} - \frac{\mcal{Q}_{T}(\sqrt{\beta/x}\mcal{M}, \Kt, Y_\mathbb{P})}{\sqrt{\beta/x}\mcal{M}}\right)^2,
\end{align}
\begin{align}
    \label{cross3_tmd}
    \frac{\rmd\sigma_{(\bar{q}q)}^
	{\gamma_{\scriptscriptstyle T}^* A\rightarrow (q)\bar q g A}}
	{\rmd z_2
	\rmd z_3
	\rmd^{2}\Pt
  	\rmd^{2}\Kt\,
  	 \rmd \ln(1/x)}
   &= -\frac{1}{(1-x)}\,\frac{1}{Q^2}\,\frac{S_{\perp} \alpha_{\rm em} N_c}{2\pi^2} 
   \left( \sum e_f^2\right)
   \mcal{H}^{(\bar{q}q)}_T
 	(z_2,z_3,P_{\perp}^2,\tilde{Q}^2)\,
   \big|\mcal{Q}_{T}(\mcal{M}, \Kt,Y_{\mathbb P})\big|\nonumber\\
   &~~~~\times \mcal{M}\left(\frac{\mcal{Q}_{T}(\mcal{M}, \Kt, Y_\mathbb{P})}{\mcal{M}} - \frac{\mcal{Q}_{T}(\sqrt{\beta/x}\mcal{M}, \Kt, Y_\mathbb{P})}{\sqrt{\beta/x}\mcal{M}}\right),
\end{align}
and
\begin{align}
    \label{cross2_tmd}
    \frac{\rmd\sigma_{(qq)}^
	{\gamma_{\scriptscriptstyle T}^* A\rightarrow (q)\bar q g A}}
	{\rmd z_2
	\rmd z_3
	\rmd^{2}\Pt
  	\rmd^{2}\Kt\,
  	 \rmd \ln(1/x)}
   &= \frac{1}{(1-x)}\,\frac{1}{Q^2}\,\frac{S_{\perp} \alpha_{\rm em} N_c}{2\pi^2} 
   \left( \sum e_f^2\right)
   \mcal{H}^{(qq)}_T
 	(z_2,z_3,P_{\perp}^2,\tilde{Q}^2)\,
   \big|\mcal{Q}_{T}(\mcal{M}, \Kt,Y_{\mathbb P})\big|^2.
\end{align}
\end{widetext}
In the soft parts of Eq. \eqref{cross1_tmd} and \eqref{cross3_tmd}, we have used the delta function and the relation $\mcal{M}^2 = x \Kt^2/(1-x)$ to write
\begin{align}
z_1
&= \frac{x}{1-x}\frac{\Kt^2}{\frac{\Pt^2}{z_2z_3} + Q^2} = \frac{\mcal{M}^2\beta}{xQ^2},
\end{align}
and therefore get
\begin{equation}
    \sqrt{z_1}Q = \sqrt{\frac{\beta}{x}}\mcal{M}\,.
\end{equation}
The hard parts $\mcal{H}_T$, which are identical to those in Ref.~\cite{Hauksson:2024bvv} (see Eqs. (4.9), (4.11) and (4.12) therein) are given by
\begin{align}
	\label{hqqg}
	\mcal{H}^{(\bar{q}\bar{q})}_T
 	(z_2,z_3,P_{\perp}^2,\tilde{Q}^2)
	\equiv 
	\delta_{z}\,
	\frac{\alpha_s}{2\pi^2}\,
	P_{gq}(z_3)\,
	\frac{\tilde{Q}^2}
	{P^2_\perp(P_{\perp}^2 + \tilde{Q}^2)},
\end{align}   
\begin{align}
    \mcal{H}^{(\bar{q}q)}_T
 	(z_2,z_3,P_{\perp}^2,\tilde{Q}^2) &= \frac{\alpha_s C_F}{\pi^2}\,\delta_{z}\,
   \frac{\tilde{Q}^2}{(P_{\perp}^2 + \tilde{Q}^2)^2}\,
   \frac{z_2^2}{z_3}
\end{align}
and
\begin{align}
	\label{hqqg2}
	\mcal{H}_T^{(qq)}
   (z_2,z_3,P_{\perp}^2,\tilde{Q}^2)
	&\equiv 
	\delta_{z}\,
	\frac{\alpha_s C_F}{\pi^2}\,P_{q\gamma}(z_2) \,
	\frac{1}{z_3}\,
	\frac{\tilde{Q}^2 P^2_{\perp}}
	{(P_{\perp}^2 + \tilde{Q}^2)^3}.
\end{align} 

We note that if it were not for the $\mcal{Q}_{T}\left(\sqrt{\frac{\beta}{x}}\mcal{M}, \Kt, Y_\mathbb{P}\right)/\left(\sqrt{\frac{\beta}{x}}\mcal{M}\right)$ term occuring in Eqs.~\eqref{cross1_tmd} and \eqref{cross3_tmd}, TMD factorization would be manifest at this stage with the definition of the quark TMD being~\cite{Hauksson:2024bvv},
\begin{align}
  \label{qDTMD}
  \frac{\rmd x q_{\mathbb{P}}
  (x, x_{\mathbb{P}}, \Kt^2)}{\rmd^2 \Kt}\equiv
  \frac{S_{\perp} N_c}{4 \pi^3}\,	
	\frac{[\mcal{Q}_T(\mcal{M}, \Kt, Y_\mathbb{P})]^2}
	{2\pi(1-x)}.
\end{align}
For notational convenience we define the following two TMD-like objects, which we must caution {\it are not in fact TMDs} since they do not result in factorization:
\begin{align}
  \label{mod_qDTMD_1}
  \frac{\rmd x\tilde{q}_{\mathbb{P}}
  (x, x_{\mathbb{P}}, \Kt^2)}{\rmd^2 \Kt}&\equiv
  \frac{S_{\perp} N_c}{4 \pi^3}\,	
	\frac{\mcal{M}^2}
	{2\pi(1-x)}\nonumber \\ &\hspace*{-1.6cm}\times\left(\frac{\mcal{Q}_{T}(\mcal{M}, \Kt, Y_\mathbb{P})}{\mcal{M}} - \frac{\mcal{Q}_{T}\left(\sqrt{\beta/x}\mcal{M}, \Kt, Y_\mathbb{P}\right)}{\sqrt{\beta/x}\mcal{M}} \right)^2,
\end{align}
\begin{align}
  \label{mod_qDTMD_2}
  \frac{\rmd x\hat{q}_{\mathbb{P}}
  (x, x_{\mathbb{P}}, \Kt^2)}{\rmd^2 \Kt}&\equiv
  \frac{S_{\perp} N_c}{4 \pi^3}\,	
	\frac{\mcal{M}}
	{2\pi(1-x)}\mcal{Q}_{T}(\mcal{M}, \Kt, Y_\mathbb{P})\nonumber\\
    &\hspace*{-1.6cm}\times\left(\frac{\mcal{Q}_{T}(\mcal{M}, \Kt, Y_\mathbb{P})}{\mcal{M}} - \frac{\mcal{Q}_{T}\left(\sqrt{\beta/x}\mcal{M}, \Kt, Y_\mathbb{P}\right)}{\sqrt{\beta/x}\mcal{M}} \right).
\end{align}

In terms of the quark DTMD defined in Eq.~\eqref{qDTMD} and the TMD-like objects defined in Eqs.~\eqref{mod_qDTMD_1} and \eqref{mod_qDTMD_2}, we can write the contributions to the differential cross section from the direct antiquark, interference and direct quark terms as,
\begin{widetext}
\begin{align}
	\label{cross1new}
	\frac{\rmd\sigma_{(\bar{q}\bar{q})}^
	{\gamma_{\scriptscriptstyle T}^* A\rightarrow (q)\bar q g A}}
	{\rmd z_2
	\rmd z_3
	\rmd^{2}\Pt
  	\rmd^{2}\Kt\,
  	 \rmd \ln(1/x)}
   = 
   \frac{4 \pi^2 \alpha_{\rm em}}{Q^2}\,  \left( \sum e_f^2\right)\,
   \mcal{H}_T^{(\bar{q}\bar{q})}
   (z_2,z_3,P_{\perp}^2,\tilde{Q}^2)\,
   \frac{\rmd x\tilde{q}_{\mathbb{P}}
  (x, x_{\mathbb{P}}, \Kt^2)}
  {\rmd^2 \Kt},
\end{align}
\begin{align}
	\label{cross3new}
	\frac{\rmd\sigma_{(\bar{q}q)}^
	{\gamma_{\scriptscriptstyle T}^* A\rightarrow (q)\bar q g A}}
	{\rmd z_2
	\rmd z_3
	\rmd^{2}\Pt
  	\rmd^{2}\Kt\,
  	 \rmd \ln(1/x)}
   = 
   \frac{4 \pi^2 \alpha_{\rm em}}{Q^2}\,  \left( \sum e_f^2\right)\,
   \mcal{H}_T^{(\bar{q}q)}
   (z_2,z_3,P_{\perp}^2,\tilde{Q}^2)\,
   \frac{\rmd x\hat{q}_{\mathbb{P}}
  (x, x_{\mathbb{P}}, \Kt^2)}
  {\rmd^2 \Kt},
\end{align}
\begin{align}
	\label{cross2new}
	\frac{\rmd\sigma_{(qq)}^
	{\gamma_{\scriptscriptstyle T}^* A\rightarrow (q)\bar q g A}}
	{\rmd z_2
	\rmd z_3
	\rmd^{2}\Pt
  	\rmd^{2}\Kt\,
  	 \rmd \ln(1/x)}
   = 
   \frac{4 \pi^2 \alpha_{\rm em}}{Q^2}\,  \left( \sum e_f^2\right)\,
   \mcal{H}_T^{(qq)}
   (z_2,z_3,P_{\perp}^2,\tilde{Q}^2)\,
   \frac{\rmd xq_{\mathbb{P}}
  (x, x_{\mathbb{P}}, \Kt^2)}
  {\rmd^2 \Kt}.
\end{align}
\end{widetext}

At this stage we have the fully differential cross section for the diffractive production of a $q\bar{q}g$ final state with a soft quark, where the gluon emission occurs before the shockwave. Ultimately we are interested in the corresponding contribution to the diffractive structure function $F^D_{T,q\bar{q}g}$. To obtain this, one has to integrate over the kinematical variables $x$, $z_2$, $z_3$, $\Pt$ and $\Kt$, while keeping the invariant mass or, equivalently $\beta$, fixed. Given that $\beta$ and the momentum fraction carried by the Pomeron $\xpom$ are related to Bjorken-$x$ as $x_{\mathrm{Bj}} = \beta\xpom$, this is equivalent to holding the rapidity gap $Y_\mathbb{P}=\ln(1/\xpom)$ fixed. 

As an intermediate step, we first obtain the diffractive SIDIS cross section. This would correspond to only one of the hard jets in the dijet (could be either one of them) being measured in the final state. To get the diffractive SIDIS cross section, we integrate over all the variables that are not observed in the final state, i.e., $x$, $z_2$, $z_3$, and $\Kt$. With this procedure we get for the direct antiquark term,
\begin{multline}
       \label{SIDIS_direct_qbar}
	\frac{\rmd \sigma_{(\bar{q}\bar{q})}^{\gamma_{\scriptscriptstyle T}^* A\rightarrow (q)\bar q g A}}{ \rmd^{2}\Pt \,\rmd \ln(1/\beta)} 
	=
      \int \rmd z_2\rmd z_3
  	  \int \frac{\rmd x}{x}\,\beta\,\delta\left(\beta - x\frac{\tilde{Q}^2}{\tilde{Q}^2 + \Pt^2}\right) \\
   \times \int \dd[2]{\Kt} \frac{\rmd\sigma_{(\bar{q}\bar{q})}^
	{\gamma_{\scriptscriptstyle T}^* A\rightarrow (q)\bar q g A}}
	{\dd{z_2}
	\dd{z_3}
	\dd[2]{\Pt}
  	\dd[2]{\Kt}
  	 \dd{ \ln(1/x)}}\\
  =  
   \frac{4\pi^2\alpha_{\rm em}}{Q^2}\,\left( \sum e_f^2\right)\int \dd{z_2} \dd{z_3}
  	  \int_\beta^1\dd{x} \delta\left(x-\beta\, \frac{\Pt^2+\tilde{Q}^2}{\tilde{Q}^2}\right) \\
    \hspace{-2.5cm}\times  \,
   \,\mcal{H}^{(\bar{q}\bar{q})}_T
 	(z_2,z_3,P_{\perp}^2,\tilde{Q}^2)\, x\tilde{q}_{\mathbb{P}}
  (x, x_{\mathbb{P}}, (1-x)P_{\perp}^2),
\end{multline}
where in the last line, we have integrated the TMD-like object over $\Kt$ to get a PDF-like object. The resolution scale $(1-x)\Pt^2$ which is essentially the upper limit of the $\Kt^2$ in the integration, has been chosen based on the arguments presented in Section 4 of Ref.~\cite{Hauksson:2024bvv}. If we define,
\begin{align}
    H^{(\bar{q}\bar{q})}_T(z_2,z_3,P_{\perp}^2,\tilde{Q}^2) \equiv \frac{4\pi^2\alpha_{\rm em}}{Q^2}\left( \sum e_f^2\right)\,\mcal{H}^{(\bar{q}\bar{q})}_T,
\end{align}
we get an expression looking very similar to Eq. (5.1) of Ref.~\cite{Hauksson:2024bvv},
\begin{multline}
\label{Eq51}
	\frac{\rmd \sigma^{\gamma_{\scriptscriptstyle T}^* A\rightarrow (q)\bar q g A}}{ \dd[2]{\Pt} \dd{ \ln(1/\beta)}} 
       =	
       \int \dd{z_2}\dd{z_3}
  	  \int_\beta^1\dd x\,\delta\left(x-\beta\, \frac{\Pt^2+\tilde{Q}^2}{\tilde{Q}^2}\right)\\
      \times   	H^{(\bar{q}\bar{q})}_T(z_2,z_3,P_{\perp}^2,\tilde{Q}^2)\,
  	x\tilde{q}_{\mathbb{P}}\left(x, x_{\mathbb{P}}, (1-x)\Pt^2\right).	  
\end{multline}

The integration over $z_2$ and $z_3$ can be performed using the delta functions $\delta(1-z_2-z_3)$ (which is implicit in  $H^{(\bar{q}\bar{q})}_T$) and $\delta\left(x-\beta\, \frac{\Pt^2+\tilde{Q}^2}{\tilde{Q}^2}\right)$ (recall that $\tilde{Q}^2=z_2z_3Q^2$). As shown in Ref.~\cite{Hauksson:2024bvv} (see discussion around Eq. (5.2)), this implies a lower bound on $x$ that is slightly larger than $\beta$, $x_\text{min} = \beta(1+4\Pt^2/Q^2)$. This turns out to be important in their calculation as the integrand contains singularities at $x=\beta$. Specifically, as shown in Appendix D therein, the hard part of the direct term from gluon emission by the antiquark contains both logarithmic and power-like divergences as $x\to\beta$. However in our case, the soft part also depends on $x-\beta$ (which is not the case in Ref. \cite{Hauksson:2024bvv}) and this dependence cancels out the singularities in the hard part as we will shortly see. 

To proceed let us write $H^{(\bar{q}\bar{q})}_T$ as,
\begin{align}
    \label{HT}
    H^{(\bar{q}\bar{q})}_T(z_2,z_3,P_{\perp}^2,\tilde{Q}^2)\,&=\,
\frac{4 \pi^2 \alpha_{\rm em}}{Q^2} \left( \sum e_f^2\right)
   	\delta_{z}\nonumber\\
    &\times 	\frac{\alpha_s C_F}{2\pi^2}\,\frac{1}{P^2_\perp}\,
	\frac{1}{z_3}
	h_T^{(\bar q)}
	\,,
\end{align}
where,
\begin{align}
    h_T^{(\bar q)} =\,\frac{(1+z_2^2)\tilde{Q}^2}
	{P_{\perp}^2 + \tilde{Q}^2}\,.
\end{align}
The hard factor $h_T^{(\bar q)}$ is identical to that in Ref.~\cite{Hauksson:2024bvv} (c.f. Eq. (D.1)) and we can perform the integration over $z_2$ and $z_3$ exactly as shown in their Appendix D. Namely, combining the factor $1/z_3$ in Eq. \eqref{HT} with the delta function in Eq. \eqref{Eq51}, we have,
\begin{multline}\label{delta3}
\frac{1}{z_3}\delta\left(x-\beta\, \frac{\Pt^2+\tilde{Q}^2}{\tilde{Q}^2}\right)
=\,\frac{z_2}{x-\beta}\,\delta\left(z_2z_3-\frac{\beta}{x-\beta}\,\frac{\Pt^2}{Q^2}\right)
\\
=\,\frac{z_2}{x-\beta}\,
 \frac{\delta\left(z_2-z_*\right) + \delta\left(z_2-1+z_*\right)}{1-2z_*}.
   \end{multline}
Here $z_*$ is the solution smaller than 1/2:
\begin{equation}
	\label{xmin}
	z_*\,=\,\frac{1}{2}\left(1 -\sqrt{1-\frac{4\beta}{x-\beta}\frac{\Pt^2}{Q^2}}\right)\,.
\end{equation}
Using this to perform the $z_2$ and $z_3$ integrals we have,
\begin{multline}
\label{vartheta_integral}
\int \dd{z_2}
  	\dd{z_3} \delta_{z}\,\delta\left(x-\beta\frac{\Pt^2+\tilde{Q}^2}{\tilde{Q}^2}\right)\frac{h_T^{(\bar q)}}{z_3} \times \text{(soft part)}  \\
		=\left(\frac{\beta}{x(x-\beta)}\right)\left[2-\frac{3\beta}{x-\beta}\frac{\Pt^2}{Q^2}\right]\frac{1}{1-2z_*}\\
        \times  \frac{\mcal{M}^2}
	{1-x}\left(\frac{\mcal{Q}_{T}(\mcal{M}, \Kt, Y_\mathbb{P})}{\mcal{M}} - \frac{\mcal{Q}_{T}\left(\sqrt{\beta/x}\mcal{M}, \Kt, Y_\mathbb{P}\right)}{\sqrt{\beta/x}\mcal{M}} \right)^2.
\end{multline}

Now we consider the singular structure of the integral over $x$ involving this contribution. Ignoring irrelevant factors such as $1/x$ and $1/(1-x)$, we have an integral of the form,
\begin{equation}
\int_{x_{\rm min} }^1\frac{\rmd x}{x-\beta}\,\frac{2-\frac{3\beta}{x-\beta}\frac{\Pt^2}{Q^2}}
{\sqrt{1-\frac{4\beta}{x-\beta}\frac{\Pt^2}{Q^2}}}\times \text{(soft part)}
\end{equation}
In the leading twist approximation we can ignore terms proportional to $\Pt^2/Q^2$ or higher powers of it. This would also entail replacing the lower bound $x_\text{min}$ with $\beta$. In this case, the hard part clearly diverges as $x\to\beta$. However this potential divergence is cancelled by the soft part. Using a change of variables $t \equiv  x - \beta$ we have
\begin{multline}
\label{soft_and_hard_expanded}
\int_{x_{\rm min} }^1 \frac{\rmd x}{x-\beta}\,\frac{2-\frac{3\beta}{x-\beta}\frac{\Pt^2}{Q^2}}
{\sqrt{1-\frac{4\beta}{x-\beta}\frac{\Pt^2}{Q^2}}}\times \text{(soft part)}\\
=\int_{t_{\rm min} }^{1-\beta}\frac{\rmd t}{t}\left\{2-\frac{t_{\rm min}}{4t}+\,\order{(t_{\rm min}/t)^2}\right\}\\
\times (S_1~t  + \mcal{O}(t^2))^2,
\end{multline}
where $t_\text{min} = 4 \beta \Pt^2/Q^2$ and on the RHS, we expanded both the hard part and the soft part in $t$. Here $S_1$ represents the linear term in the expansion of the factor $(\mcal{Q}_{T}(\mcal{M}, \Kt)/\mcal{M} - \mcal{Q}_{T}(\sqrt{\beta/x}\mcal{M}, \Kt)/\sqrt{\beta/x}\mcal{M})$. It is clear from the form of the soft part that it vanishes when $t\to 0$ since the two terms cancel each other when $x\to\beta$. The difference of the two $\mcal{Q}_T$ terms in the soft part admits an expansion in $t$ with a non-zero linear term. The leading contribution in the expansion of the soft part is thus proportional to $t^2$. This combines with the $1/t$ from the hard part and ensures that there is no singularity in the integral, i.e., all terms in the integrand are of the form $t^n$, where $n\geq 1$, and hence give a finite contribution. However, the higher order terms in the expansion of the hard part result in integrands suppressed by powers of $t_{\rm min}\sim \Pt^2/Q^2$ and thus, correspond to higher twist. Therefore we can conclude that only the leading term in the expansion of the hard part and all the terms of the soft part contribute at leading twist. 

The above analysis allows us to replace the RHS of \eqref{vartheta_integral} with,
\begin{multline}
    \frac{2\beta}{x(x-\beta)}\times  \frac{\mcal{M}^2}
	{1-x}\left(
\vphantom{\frac{\mcal{Q}_{T}\left(\sqrt{\beta/x}\mcal{M}, \Kt, Y_\mathbb{P}\right)} 
    {\sqrt{\beta/x}\mcal{M}}}  \frac{\mcal{Q}_{T}(\mcal{M}, \Kt, Y_\mathbb{P})}{\mcal{M}}  \right. \\
    - \left.\frac{\mcal{Q}_{T}\left(\sqrt{\beta/x}\mcal{M}, \Kt, Y_\mathbb{P}\right)}{\sqrt{\beta/x}\mcal{M}} \right)^2.
\end{multline}
With this we get the contribution to SIDIS from the direct term from the gluon emission by the antiquark:
\begin{align}
    \label{SIDIS_direct_qbar_final}
    \frac{\rmd \sigma_{(\bar{q}\bar{q})}^{\gamma_{\scriptscriptstyle T}^* A\rightarrow (q)\bar q g A}}{ \rmd^{2}\Pt \,\rmd \ln(1/\beta)} 
	&\,=\frac{4 \pi^2 \alpha_{\rm em}}{Q^2} \left( \sum e_f^2\right)\frac{\alpha_s C_F}{2\pi^2}\frac{1}{P^2_\perp}\nonumber\\
    &\times\int_\beta^1\rmd x\,\frac{2\beta}{x(x - \beta)}
	\,
  	x\tilde{q}_{\mathbb{P}}\left(x, x_{\mathbb{P}}, (1-x)\Pt^2\right).
\end{align}

Following a similar procedure, we get the contribution to SIDIS from the interference of gluon emissions by quark and antiquark:
\begin{align}
\label{SIDIS_direct_interf_recalc}
    \frac{\rmd \sigma_{(\bar{q}q)}^{\gamma_{\scriptscriptstyle T}^* A\rightarrow (q)\bar q g A}}{ \rmd^{2}\Pt \,\rmd \ln(1/\beta)} 
	&\,=\frac{4 \pi^2 \alpha_{\rm em}}{Q^2} \left( \sum e_f^2\right)\frac{\alpha_s C_F}{2\pi^2}\frac{1}{P^2_\perp}\nonumber\\
    &\times\int_\beta^1\rmd x\,\left(-\frac{2\beta}{x^2}\right)
	\,
  	x\hat{q}_{\mathbb{P}}\left(x, x_{\mathbb{P}}, (1-x)\Pt^2\right).
\end{align}
The direct term from gluon emission by the quark remains unchanged from Ref.~\cite{Hauksson:2024bvv}.

With these we can finally obtain the soft quark contribution to the structure function by integrating over the transverse momentum of the observed jet:
\begin{widetext}
\begin{align}
\label{eq:softq_structurefun}
    x_{\mathbb P}F_{T,q\Bar{q}g}^\textrm{D (soft-q)}(\xpom, \beta, Q^2)  &=\frac{\alpha_s C_F S_T N_c}{8\pi^4}\,\left(\sum_f e_f^2\right)\int_0^{Q^2} \rmd k^2\,\log\left(\frac{Q^2}{k^2}\right)\,\int_\beta^1\rmd x\nonumber\\
    &\,\,\,\,\,\,\times\Biggl\{\frac{2\beta}{x(x-\beta)}
	\,
	\left[\mcal{Q}_T\left(\sqrt{x}k, \sqrt{1-x}k, Y_\mathbb{P}\right) - \sqrt{\frac{x}{\beta}}\mcal{Q}_T\left(\sqrt{\beta}k, \sqrt{1-x}k, Y_\mathbb{P}\right)\right]^2
	\nonumber\\
    &\,\,\,\,\,\,+ \frac{\beta(x - \beta)}{x^3}
	\,
	\left[\mcal{Q}_T\left(\sqrt{x}k, \sqrt{1-x}k, Y_\mathbb{P}\right)\right]^2
	\nonumber\\
    &\,\,\,\,\,\,+\left(-\frac{2\beta}{x^2}\right)
	\,
	\mcal{Q}_T\left(\sqrt{x}k, \sqrt{1-x}k, Y_\mathbb{P}\right)\,\left[\mcal{Q}_T\left(\sqrt{x}k, \sqrt{1-x}k, Y_\mathbb{P}\right) - \sqrt{\frac{x}{\beta}}\mcal{Q}_T\left(\sqrt{\beta}k, \sqrt{1-x}k, Y_\mathbb{P}\right)\right]\Biggr\}.
\end{align}
Here we have also included the soft antiquark contribution, which is identical to the soft quark one. This is the soft quark counterpart to the ``W\"usthoff'' result discussed in Sec.~\Ref{sec:softg}.

\end{widetext}

\subsection{Large $M_X^2$ limit}
\label{sec:MSlimit}
As already discussed above, the so called ``Munier--Shoshi'' result~\cite{Munier:2003zb} corresponds to the limit of a very soft gluon ($z_3\ll 1$) 
and a very large final state mass ($M_X^2\gg Q^2$, i.e., $\beta\to 0$ limit). The cross section can be derived by including both gluon emission before and after the shockwave. The resulting cross section given e.g. in Refs.~\cite{Kowalski:2008sa,Beuf:2022kyp}, in the case of a large uniform nucleus, reads
\begin{multline}
\label{eq:mslimit}
    \xpom F_{T,q\Bar q g}^{\textrm{D}\ (\text{MS})}(\xpom, \beta=0, Q^2)
    = 
    \frac{\as N_c C_F S_T Q^2}{16\pi^5 \aem} 
    \int \dd[2]{\rt} \dd[2]{\rt'}  \\
    \times \int_{0}^1 \frac{\dd z}{z(1-z)} 
    \left |\widetilde{\psi}_{\gamma^{*}_T \rightarrow q\bar q}^{\rm LO}(\rt,z) \right|^2  \frac{\rt^2}{\rt'^2 (\rt-\rt')^2} 
    \\ \times
    \bigg[ N(\rt') + N(\rt-\rt') - N(\rt) - N(\rt')N(\rt-\rt') \bigg]^2.
\end{multline}
Here  $\widetilde{\psi}_{\gamma^*_T \to q\bar q}^\mathrm{LO}$ is the leading order transverse virtual photon wave function and is given 
by,
\begin{multline}
    \widetilde{\psi}_{\gamma^*_T \to q\bar q}^\mathrm{LO}(\rt,z) = 2\left(\sum_f e_f^2\right)\frac{\alpha_\text{em}}{\pi}~z^2(1-z)^2\\\left(z^2 + (1-z)^2\right)Q^2~\besk_{1}^2(|\rt|\Bar{Q})\,,
\end{multline}
where $\Bar{Q}\equiv\sqrt{z(1-z)}Q$.
In Appendix~\ref{appendix:MS} we demonstrate how, following Ref.~\cite{Beuf:2022kyp}, this limit can be recovered from the NLO-trip term~\eqref{eq:trip} by modifying the Wilson line structure.

\section{Numerical results}
\label{sec:numerics}

We begin the numerical analysis by determining the accuracy of the existing approximative results for the diffractive  $q\bar q g$ production. We numerically evaluate the ``NLO-trip'' contribution to the diffractive cross section in the case of a large and uniform target as discussed in Sec.~\ref{sec:nlo_qqg}. The obtained $q\bar q g$ contribution to the diffractive structure function is then compared to the results obtained in the soft gluon and soft quark limits discussed in Sec.~\ref{sec:limits}. 

Our numerical implementation to evaluate the cross section \eqref{eq:trip}, written in Julia, is publicly available~\cite{zenodorepo}. We have used a fixed coupling constant $\alpha_s = 0.2$. In setting up the integral, we make a change of variables from the transverse coordinates of the partons $\xt_i$, $\xt_{\ov i}$ to the vectors $\xt_{31}$, $\xt_{32}$, $\xt_{\ov 3\ov 1}$, $\xt_{\ov 3\ov 2}$, $\bt$ and $\ov \bt$. Note that since we only consider the cross-section integrated over the momentum transfer $\Deltat$ and assume that the impact parameter dependence factorizes out from the dipole, we can use Eq.~\eqref{eq:factorized_Delta_integral} to perform the integrals over $\bt$ and $\ov \bt$ analytically. The integration over the remaining transverse separations is performed in polar coordinates. The assumption of a factorized impact parameter dependence means that the dipole correlators are independent of the dipole orientation. This means that there is a global rotational symmetry which we use to fix the azimuthal angle of $\xt_{31}$, $\phi_{31} = 0$. The magnitudes of the transverse vectors $|\xt_{ij}|$ are integrated from 0 to $x_{\text{max}} = 40~\text{GeV}^{-1}$. We have 
found that good convergence is achieved already around $x_{\text{max}}= 20~\text{GeV}^{-1}$. In the end, after using the delta function in Eq.~\eqref{eq:phase_space} to perform the integral over $z_3$, we are left with a 9D numerical integral. This computationally challenging integral is evaluated using the VEGAS algorithm implemented in the MCIntegration package~\cite{MCIntegration_jl}, which supports parallel evaluation using the Message Passing Interface (MPI) framework. Convergence is slower for larger values of $Q^2$ ($\gtrsim 50~\text{GeV}^2$) and/or very small or very large values of $\beta$ but even in such cases, using $4\times 10^{10}$ integration points results in a relative error of around 1\% or less. Such an integration typically takes 30 hours when parallelized over 16 cores on an Intel Xeon Gold 6230 processor in a standard compute node of the Puhti cluster at CSC -- IT Centre for Science in Finland.

\begin{figure}[tb]
    \centering
    \includegraphics[width=\columnwidth]{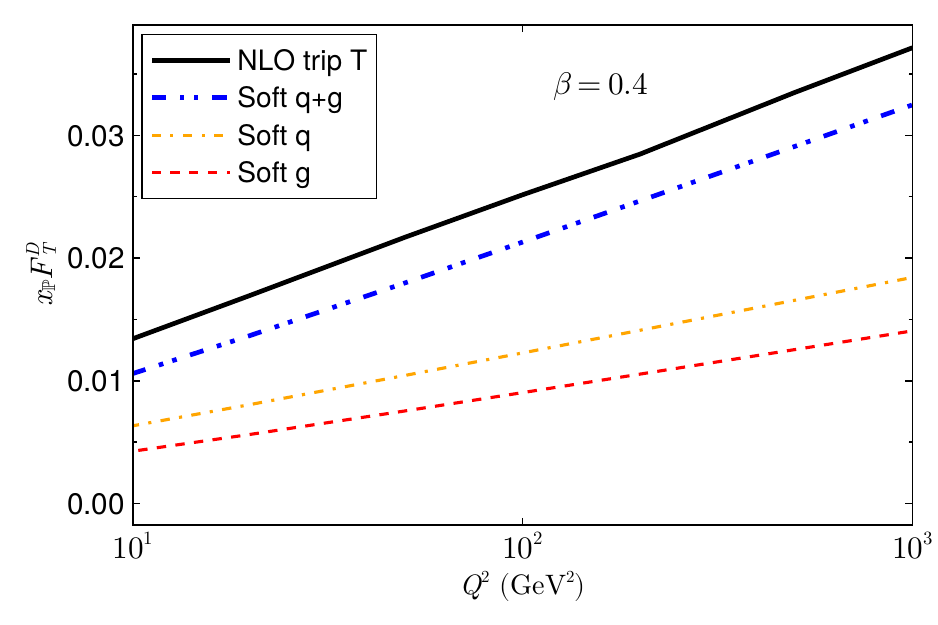}
    \caption{
    Contribution to the diffractive structure function from the $q\bar qg $ production channel. 
     }
    \label{fig:q2dep_separately}
\end{figure}

As only the transverse photon contributes at high-$Q^2$, we first focus on the transverse structure function $\xpom F_{T}^{\mathrm{D}}$. Fig.~\ref{fig:q2dep_separately} shows the diffractive structure function as a function of $Q^2$ calculated by evaluating the production cross section in exact kinematics (NLO-trip, Eq.~\eqref{eq:trip}) and by separately computing the soft quark~\eqref{eq:softq_structurefun}, and soft gluon~\eqref{eq:wusthoffqqbarg} contributions. 
The results are shown at $\beta=0.4$ which, as we will show explicitly soon, approximately corresponds to kinematics where the $q\bar qg $ production channel becomes numerically important.
In particular we notice that in the considered kinematical domain, relevant for the future EIC as well as higher-energy LHeC/FCC-he, the soft gluon contribution alone is not a good approximation for the diffractive $q\bar q g$ production.  

The soft quark production channel is found to be comparable to, and even somewhat larger, than the soft gluon contribution. This is because, although soft gluon emission is favored in QCD, destructive interference between the gluon emissions from the quark and the antiquark  suppresses this channel, see Ref.~\cite{Hauksson:2024bvv} for a more detailed discussion. At high $Q^2$, the combined soft quark+gluon contribution provides a relatively good estimate for the full NLO-trip contribution.

\begin{figure}[tb]
    \centering
    \includegraphics[width=\columnwidth]{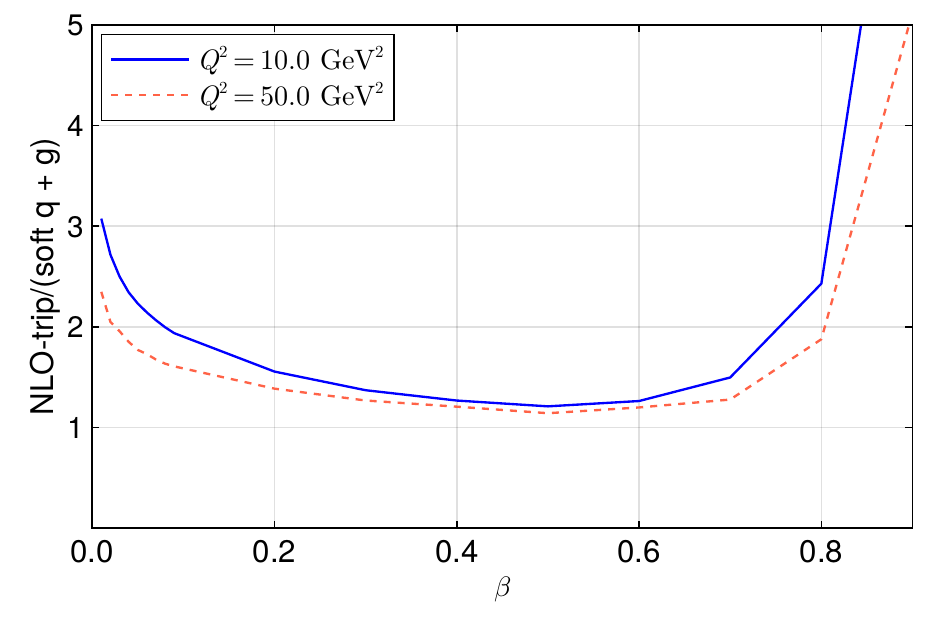}
    \caption{Relative accuracy of estimating the NLO-trip contribution by the soft quark and soft gluon contributions. }
    \label{fig:largeq2lim_softq_plus_softg}
\end{figure}

Next we quantify in more detail how well the NLO-trip contribution can be estimated by the combined soft quark and  gluon terms. Fig.~\Ref{fig:largeq2lim_softq_plus_softg} shows the NLO-trip result normalized by the sum of the soft quark and soft gluon contributions. The combined soft quark + gluon contribution  approximates the full result relatively well around $\beta \sim 0.3\dots 0.6$, with the approximation becoming better towards higher $Q^2$ corresponding to the kinematical domain where they have been derived. However, even at $Q^2=50\,\mathrm{GeV}^2$  the accuracy of this estimate is never better than $\sim 10\%$.
This is a significantly larger difference than the accuracy estimated for the diffractive structure function measurements at the EIC~\cite{AbdulKhalek:2021gbh}.

At both small and large $\beta$ the approximate results obtained in the aligned jet limit break down, as expected. The reason for this can be understood as follows. Let us first consider the large $\beta$ case. In both the soft quark and soft gluon limits, one parton carries a small fraction of the photon's plus momentum, which naturally leads to a large invariant mass $M_X^2$ in the final state. In contrast, large $\beta$ fixes the invariant mass to be small, and it therefore becomes necessary to treat  the $q\bar qg $ system kinematics exactly. 
At small $\beta$ the high-$Q^2$ limit employed in the derivation of the soft parton contributions is also not justified, as in this kinematics $M_X^2 \gg Q^2$.

\begin{figure}[tb]
    \subfloat[$Q^2=10\,\mathrm{GeV}^2$]{%
    \includegraphics[width=0.49\textwidth]{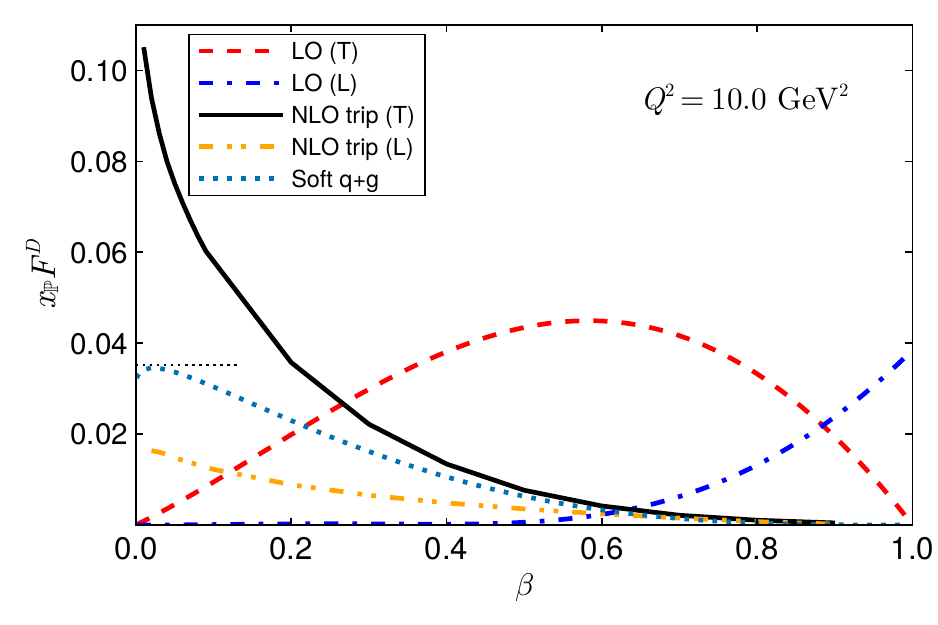}
    }
    \hfill
     \subfloat[$Q^2=50\,\mathrm{GeV}^2$]{%
    \includegraphics[width=0.49\textwidth]{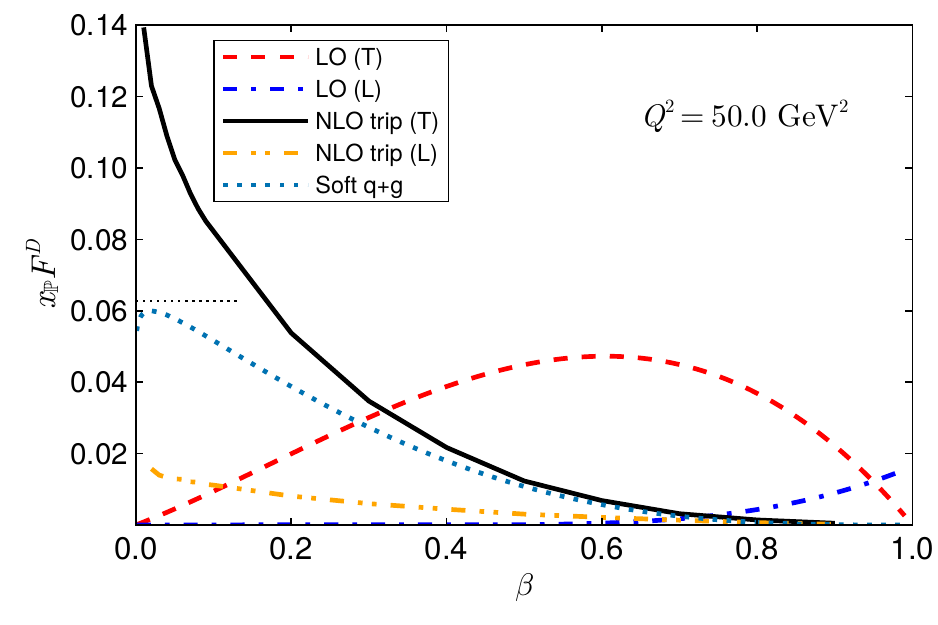}
    }
    \caption{Different contributions to the diffractive structure function as a function of $\beta$. The horizontal dotted line indicates the Munier-Shoshi limit at the given value of $Q^2$.}
    \label{fig:betadep}
\end{figure}

Next we compute different contributions to  the diffractive structure function as a function $\beta$. Now, in addition to the transverse photon contribution, we also include the longitudinal contribution to the $q\bar q g$ production, for which no estimates exist in the literature. 
Furthermore, for reference we also show the transverse and longitudinal $q\bar q$ production cross sections calculated following Refs.~\cite{Beuf:2022kyp,Kowalski:2008sa}.
The obtained dependencies on $\beta$ are shown in Fig.~\ref{fig:betadep}. 

The $q\bar q$ channel is found to dominate at large $\beta\gtrsim 0.5$. At smaller $\beta$, especially the transverse NLO-trip contribution becomes important and dominates below $\beta \lesssim 0.2$. The longitudinal photon producing the $q\bar q g$ system is comparable to the transverse production channel at lower $Q^2$ and around $\beta \gtrsim 0.6$, which is the region where the $q\bar q$ production is dominant.  Again the aligned jet limit (``Soft $q+g$'') becomes a better approximation towards higher $Q^2$, but  significant differences between the full result and the aligned jet limit can be seen in the kinematical range relevant for the EIC and FCC-he.

The ``Munier-Shoshi'' result, valid in the $\beta\to 0$ limit and discussed in Sec.~\ref{sec:MSlimit}, is also shown in Fig.~\ref{fig:betadep} as a dotted line in the small-$\beta$ region. Note that, unlike the soft quark or soft gluon contributions, this limit cannot be derived from the NLO trip term alone; one must also include the emission-after-the-shockewave contribution~\cite{Beuf:2022kyp}. This limit is found to differ significantly from the NLO trip contribution at small $\beta$. Interestingly, however, the Munier-Shoshi result is numerically close to the sum of the soft quark and soft gluon contributions.

\begin{figure}
    \centering
\includegraphics[width=\columnwidth]{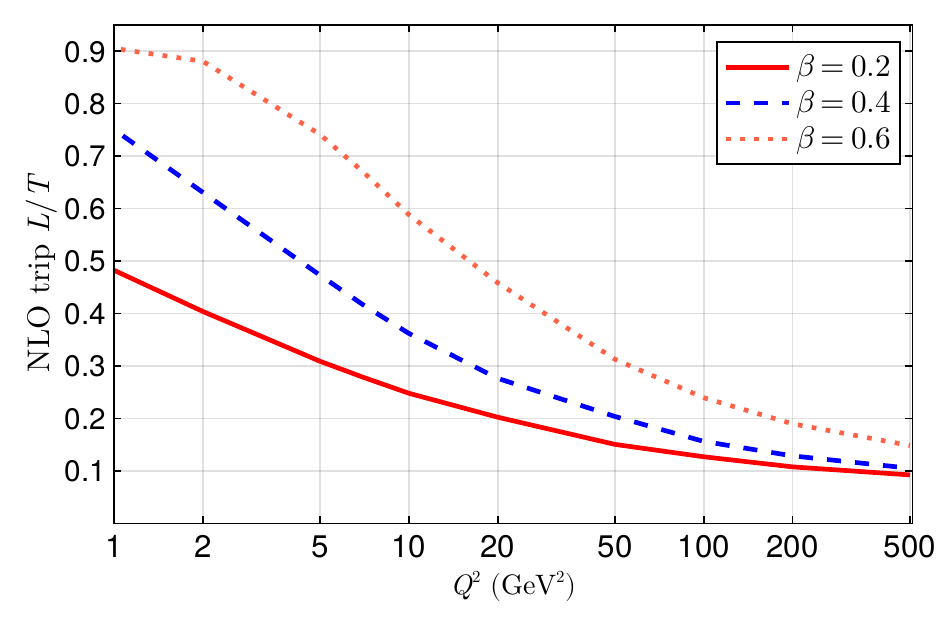}
    \caption{Ratio of the longitudinal and transverse photon components of the NLO-trip contribution.}
    \label{fig:LT_ratio}
\end{figure}

The relative importance between the transversally and longitudinally polarized $q\bar q g$ contributions is shown in Fig.~\ref{fig:LT_ratio}. The longitudinal contribution vanishes at $Q^2=0$, and does not have a $\log Q^2$ enhancement at high $Q^2$ unlike the transverse one. As such the longitudinal-to-transverse ratio vanishes both at very low and large $Q^2$. Although the transverse component dominates at high $Q^2$, in realistic EIC kinematics the longitudinal photon contribution should be included in order to achieve the percent-level precision already achieved in diffractive structure function measurements at HERA~\cite{H1:2012xlc}. Furthermore, at the EIC it will also be possible to separately measure the $F^D_L$~\cite{AbdulKhalek:2021gbh}, for which the longitudinal $q\bar qg$ component dominates below $\beta \lesssim 0.5$.


Finally we note that if one includes a subset of contributions not enhanced by large $\log Q^2$ to the aligned jet limit result, a good approximation for the full result is obtained in the kinematical domain relevant for the EIC. We demonstrate this in Appendix~\ref{appendix:lnbeta}.

\section{Conclusions}
\label{sec:conclusions}

We have calculated diffractive $q\bar q g$ production, where the three-parton system interacts coherently with the target. This process represents a finite subset of the next-to-leading order corrections to the diffractive structure functions. We have numerically evaluated  this contribution, originally derived in Refs.~\cite{Beuf:2022kyp,Beuf:2024msh}, and compared it with the previously used approximations obtained in approximative kinematics where one of the partons is soft.

We find that the commonly used ``W\"ustoff'' or ``GBW'' result, which corresponds to the soft gluon limit at high $Q^2$, does not provide an accurate estimate for the NLO contribution. Instead, it underestimates the result obtained in exact kinematics by roughly a factor of $3$ in  realistic EIC kinematics. 
We have also derived the corresponding soft quark contribution to the $q\bar q g$ production following Ref.~\cite{Hauksson:2024bvv}, including only the finite contribution in which all partons interact with the target. The result is given in Eq.~\eqref{eq:softq_structurefun}. When the soft gluon and soft quark contributions are combined, they provide a reasonable approximation to the diffractive cross section in the region $\beta \sim 0.3 \dots 0.6$. However, the deviations from the exact result are significantly larger than the  high precision expected to be achieved at the EIC. This approximation can be further improved by including a subset of contributions not enhanced by a large $\log Q^2$. 

Our findings highlight the importance of including  NLO corrections in calculations of diffractive structure functions in HERA and EIC kinematics. This is especially important because diffractive processes are expected to be especially powerful probes of gluon saturation at the EIC, where diffractive structure functions for nuclei will be measured for the first time and with high precision. In the future, we will compute diffractive structure functions for protons and nuclei consistently at the full NLO accuracy by incorporating the remaining NLO corrections derived in Ref.~\cite{Beuf:2024msh}, together with the dipole-target scattering amplitude determined at NLO accuracy~\cite{Casuga:2025etc} and evolved with the NLO BK equation~\cite{Balitsky:2008zza,Lappi:2016fmu}.

\begin{acknowledgments}
We thank H. Hänninen, M. Kampshoff, T. Lappi,  C. Royon, and F. Salazar  for discussions.
A.K and H.M are supported by the Research Council of Finland, the Centre of Excellence in Quark Matter and projects 338263 and 359902, and by the European Research Council (ERC, grant agreements ERC-2023-101123801 GlueSatLight and ERC-2018-ADG-835105 YoctoLHC).
J.P. is supported by the National Science Foundation, under grant No.~PHY-2515057, and by the U.S. Department of Energy, Office of Science, Office of Nuclear Physics, within the framework of the Saturated Glue (SURGE) Topical Theory Collaboration.
The content of this article does not reflect the official opinion of the European Union and responsibility for the information and views expressed therein lies entirely with the authors. Computing resources from CSC – IT Center for Science in Espoo, Finland and from the Finnish Computing Competence Infrastructure (persistent identifier \texttt{urn:nbn:fi:research-infras-2016072533}) were used in this work. 
\end{acknowledgments}

\appendix

\section{Aligned jet limit beyond the leading $\log Q^2$}
\label{appendix:lnbeta}
The aligned jet (soft quark or gluon) contributions discussed in Sec.~\ref{sec:limits} have been obtained at leading $\log Q^2$ accuracy. As shown in Sec.~\ref{sec:numerics}, these contributions do not provide a precise approximation to the full NLO-trip result in realistic EIC kinematics.

One possible approach to improve the accuracy of the aligned jet estimates can be found from Ref.~\cite{Beuf:2022kyp}, where the soft gluon contribution discussed in Sec.~\ref{sec:softg} has been derived from the full NLO result by explicitly taking the large-$Q^2$ limit. As a part of this calculation, one estimates
\begin{equation}
    \log \left(\frac{Q^2}{\beta k^2}\right) \approx \log \left(\frac{Q^2}{k^2}\right),
\end{equation}
which gives the logarithm seen in Eq.~\eqref{eq:wusthoffqqbarg}. As such, one straightforward approach to include (a subset of) corrections beyond the leading $\log Q^2$ accuracy is to keep the dependence on $\beta$ inside the logarithm. Analogously this can also be done for the soft quark production where the same logarithm appears.

In this Appendix, we numerically evaluate the soft quark and gluon contributions including the logarithmic dependence on $\beta$. That is, we replace both in Eqs.~\eqref{eq:wusthoffqqbarg} (soft gluon) and \eqref{eq:softq_structurefun} (soft quark) the logarithm as
\begin{equation}
\label{eq:logbeta}
    \log \left( \frac{Q^2}{k^2} \right) \to  \log \left( \frac{Q^2}{\beta k^2} \right). 
\end{equation}

\begin{figure}[tb]
    \includegraphics[width=\columnwidth]{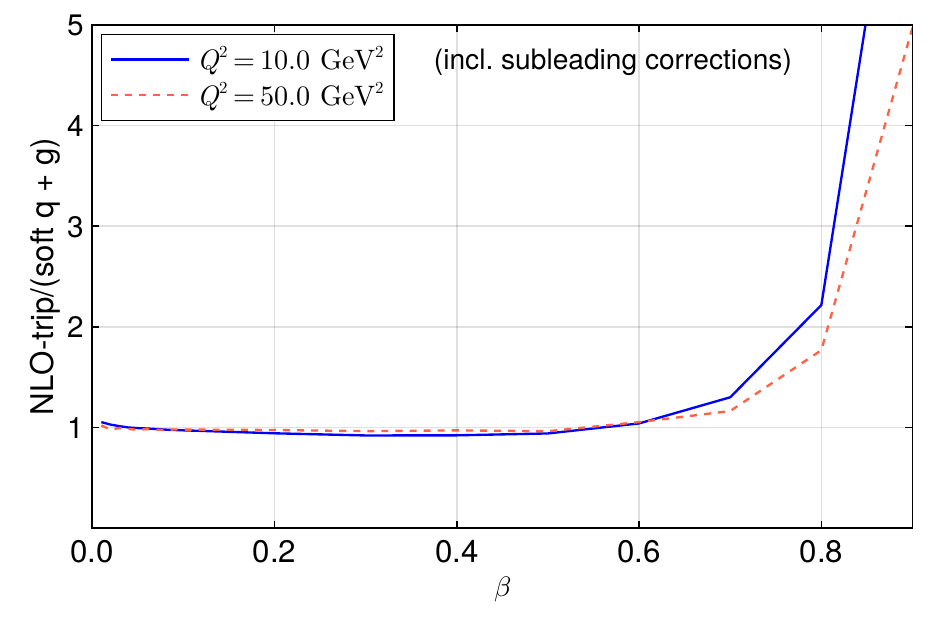}
    \caption{Relative accuracy of estimating the NLO-trip contribution by the soft quark and soft gluon contributions including the subleading $\log \beta$ dependence. }
    \label{fig:betadep_logbeta}
\end{figure}

With this modification we calculate the soft quark and gluon contributions to the diffractive $q\bar q g$ production, and compare to the exact result \eqref{eq:trip} as a function of $\beta$. A similar comparison was shown in Sec.~\ref{sec:numerics}, see Fig.~\ref{fig:betadep}, where a large difference between the aligned jet contribution and the exact result in realistic kinematics was observed, especially at $Q^2=10\,\mathrm{GeV}^2$. Results obtained using modified expressions with the $\log(\beta)$ dependence are shown in Fig.~\ref{fig:betadep_logbeta}. In this case an excellent agreement between the full and the aligned jet results is obtained even down to $Q^2=10\,\mathrm{GeV}^2$.
Over a broad range of $0.01 < \beta < 0.6$, the relative difference between the full and approximative result is at most 9\% in the studied $Q^2$ domain.
At large $\beta \gtrsim 0.7$, the relative difference between the two results remains large, but this deviation takes place in a kinematical domain where the $q\bar q$ production dominates the diffractive cross section.

We conclude that the aligned jet contribution, with the modification \eqref{eq:logbeta}, provides a good approximation for the full $q\bar q g$ production cross section (in the case where all partons interact with the target). However, it is crucial to include  both the soft quark and soft gluon contributions.

\section{Recovering the large-$M_X^2$ limit}
\label{appendix:MS}

As shown in Ref.~\cite{Beuf:2022kyp}, the ``Munier--Shoshi'' result ($\beta\to 0$ limit) can be obtained from the NLO trip result by including the corresponding emission-after-shockwave contribution. In this limit, the emission-after contribution can be included by modifying the Wilson line structure in Eq.~\eqref{eq:trip} as
\begin{multline}
\label{eq:msreplacement}
    \left(1-\tripole_{123}\right)
\left(1-\tripole_{\ov 1 \ov 2 \ov 3}\right)^\dag  
\to  \left[ \left(1-\tripole_{123}\right) - (\xt_3 \to \xt_1) ) \right] \\
\times 
\left[ \left(1-\tripole_{\ov 1 \ov 2 \ov 3 }\right) - (\xt_{\ov 3} \to \xt_{\ov 1}) ) \right]^\dag.
\end{multline}
Note that in the coincidence limit $\xt_3 \to \xt_1$ we have $\tripole_{123} \to \dipole_{12}$. 

\begin{figure}[t]
\includegraphics[width=\columnwidth]{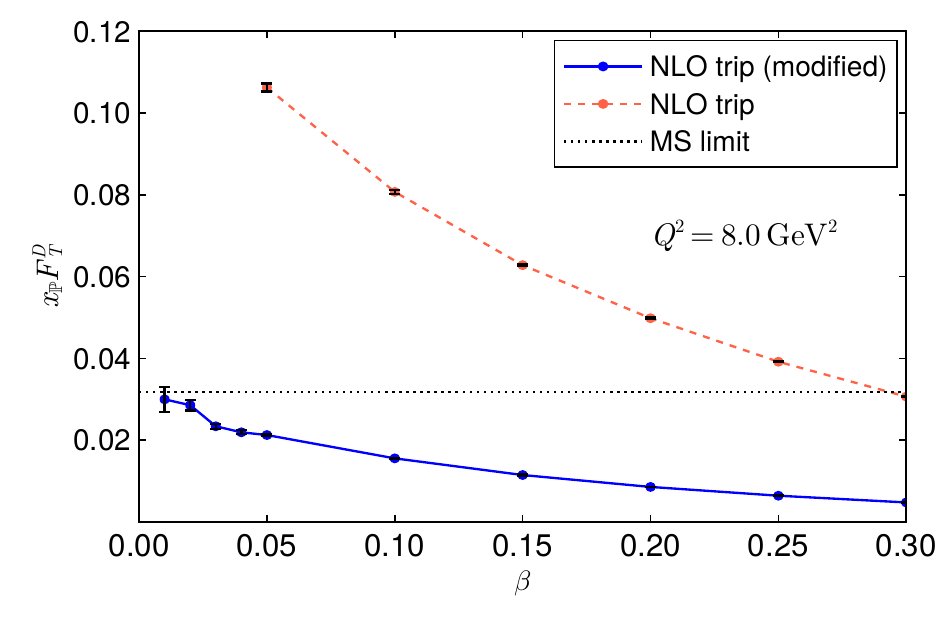}
\caption{
NLO trip contribution to the diffractive structure function, evaluated with a modified Wilson line structure corresponding to the replacement in Eq.~\eqref{eq:msreplacement} applied to Eq.~\eqref{eq:trip}, and compared to the Munier–Shoshi (MS) limit, which is $\beta$-independent. The uncertainty estimates correspond to those of the Monte Carlo integration. For reference, the NLO trip result is also shown.}
\label{fig:mscomparison}
\end{figure}

We numerically confirm the result of Ref.~\cite{Beuf:2022kyp} by computing the NLO trip contribution~\eqref{eq:trip} with a modified Wilson line structure~\eqref{eq:msreplacement} as a function of $\beta$, and comparing the result at small $\beta$ to the Munier--Shoshi limit~\eqref{eq:mslimit}. 
This comparison is shown in Fig.~\ref{fig:mscomparison}, where the diffractive structure function $F_T^D$ is shown as a function of $\beta$. 
As the Munier-Shoshi result is obtained in the $\beta\to 0$ limit, we show it as a $\beta$-independent contribution for illustrative purposes.
The numerically evaluated NLO trip contribution with a modified Wilson line structure (labeled as ``NLO trip (modified)'') is found to agree with the Munier--Shoshi limit in the small-$\beta$ region. This provides  an additional successful test for our numerical implementation.

\bibliographystyle{JHEP-2modlong}
\bibliography{refs}

\end{document}